\documentclass[useAMS,usenatbib]{mnras}

\usepackage{verbatim}
\usepackage{placeins}
\usepackage{float}
\usepackage[utf8]{inputenc}
\usepackage{mathtools}
\usepackage{amsmath}
\usepackage{amssymb}
\usepackage{graphicx}
\usepackage{multirow}
\usepackage{multicol}
\usepackage{textcomp}

\newcommand{\hthree}{\ensuremath{h_3}}
\newcommand{\hfour}{\ensuremath{h_4}}

\newcommand{\rcross}{\ensuremath{\mathrm{r_{\rm cross}}}}
\newcommand{\rbhstar}{\ensuremath{\mathrm{r_{\rm BH}^*}}}
\newcommand{\rperi}{\ensuremath{\mathrm{r_{\rm peri}}}}

\newcommand{\pibox}{\ensuremath{\pi}-box}

\title[Two phases of core formation]{The two phases of core formation - orbital evolution in the centres of ellipticals with supermassive black hole binaries }
\author[M. Frigo, T. Naab et al.]{M. Frigo$^{1,2}$\thanks{E-mail:
mfrigo@mpa-garching.mpg.de (MF)}, T. Naab$^{1}$, A. Rantala$^{1}$, P. H. Johansson$^{3}$, B. Neureiter$^{4}$, J. Thomas$^{4}$,  
\newauthor and F. Rizzuto$^{1}$\\
$^{1}$Max-Planck-Institut f{\"u}r Astrophysik, Karl-Schwarzschild-Stra{\ss}e 1, 85748 Garching, Germany\\
$^2$Excellence Cluster ORIGINS, Boltzmannstra{\ss}e 2, 85748 Garching, Germany \\
$^{3}$Department of Physics, University of Helsinki, Gustaf H{\"a}llstr{\"o}min katu 2, 00014 Helsinki, Finland\\
$^{4}$Max-Planck-Institut f{\"u}r extraterrestrische Physik, Gie{\ss}enbachstra{\ss}e 1, 85748 Garching, Germany}

\begin{document}

\date{Accepted 2021 September 18. Received 2021 September 17; in original form 2021 May 25}

\pagerange{\pageref{firstpage}--\pageref{lastpage}} \pubyear{2021}

\maketitle

\label{firstpage}

\begin{abstract}
The flat stellar density cores of massive elliptical galaxies form rapidly due to sinking supermassive black holes (SMBH) in gas-poor galaxy mergers. After the SMBHs form a bound binary, gravitational slingshot interactions with nearby stars drive the core regions towards a tangentially biased stellar velocity distribution.  We use collisionless galaxy merger simulations with accurate collisional orbit integration around the central SMBHs to demonstrate that the removal of stars from the centre by slingshot kicks accounts for the entire change in velocity anisotropy. The rate of strong (unbinding) kicks is constant over several hundred Myr at $\sim 3 \ M_\odot \rm yr^{-1}$ for our most massive SMBH binary ($M_{\rm BH} = 1.7 \times 10^{10} M_\odot$). 
Using a frequency-based orbit classification scheme (box, x-tube, z-tube, rosette) we demonstrate that slingshot kicks mostly affect box orbits with small pericentre distances, leading to a velocity anisotropy of $\beta \lesssim -0.6$ within several hundred Myr as observed in massive ellipticals with large cores. We show how different SMBH masses affect the orbital structure of the merger remnants and present a kinematic tomography connecting orbit families to integral field kinematic features. Our direct orbit classification agrees remarkably well with a modern triaxial Schwarzschild analysis applied to simulated mock kinematic maps. 
\end{abstract}

 
\section{Introduction}
Many properties of massive early-type galaxies correlate with the mass of their central super-massive black holes (SMBH), most notably galactic bulge mass and stellar velocity dispersion \citep{Dressler1989, Gebhardt2000}. This suggests a connection between SMBHs and their massive host galaxies, however, one that is not yet completely understood \citep{KormendyHo2013, Saglia2016}.
The most massive and luminous ($M_B < -20.5$) early-type galaxies (ETGs) in the Universe are particularly interesting. They have stellar masses of order $10^{12} \, M_\odot$ or larger, typically show flat stellar density profiles in their cores and host the most massive SMBHs that have been observed. The measured SMBH masses are of order $10^9 \, M_\odot$ or in some cases even $10^{10} \, M_\odot$  \citep[e.g.][]{McConnell2011, KormendyHo2013,Thomas2016, Mehrgan2019}. 

Stellar kinematics in most of these very old galaxies show little net rotation at high velocity dispersions \citep{Emsellem2007, Cappellari2016, Ene2020}. In general, photometric, stellar population, and kinematic properties of massive ETGs are consistent with a `two-phase' formation process, consisting of a rapid formation at high redshift by in-situ star formation, followed by mass assembly through mergers at redshifts $z\lesssim 2$ \citep[e.g.][]{Naab2009,Oser2010, Johansson2012, Naab2014, Penoyre2017, RodriguezGomez2016,Xu2019,Moster2020}. Here, feedback from accreting SMBHs is thought to be of fundamental importance for setting the old ages of these galaxies by suppressing or terminating star formation \citep[see e.g.][for reviews]{SomervilleDave2015,NaabOstriker2017}.

SMBHs are also thought to be responsible for the observed flat stellar density distributions in the centres of massive ellipticals \citep{Lauer1985}. The ejection of stars during the sinking and coalescence of binary SMBHs in mergers of massive ETGs is a very likely formation mechanism of such stellar density cores \citep[see e.g.][]{Milosavljevic2001, Milosavljevic2003, Merritt2013}.

Several observational facts support this. Firstly, estimated `mass deficits' in the centres of early-type galaxies are of the order of the mass of the central black hole, as predicted by theory \citep{Merritt2006,Kormendy2009}. Secondly, dynamical modelling revealed that the size of the core region corresponds almost exactly to the sphere of influence of the final central black hole, which implies a very close link between SMBH dynamics and depleted stellar cores \citep{Thomas2016}. Finally, stellar orbits in the centers of massive elliptical galaxies have been found to be tangentially biased \citep[e.g.][]{Gebhardt2003,McConnell2012}. Tangential anisotropy alone does not provide direct evidence for the SMBH binary model. However, (i) the observed orbital anisotropy profiles in elliptical galaxies with depleted stellar cores are remarkably uniform from one galaxy to another, pointing to a universal formation process of the cores. And (ii), the overall strength of the observed tangential anisotropy and its variation with radius over the core region match very well with the central structure that emerges in N-body simulations with SMBH binaries \citep{Thomas2014, Rantala2018}.
Another possible stellar core formation mechanism by SMBHs is the ejection of large amounts of nuclear gas due to active galactic nucleus (AGN) feedback, and the resulting expansion of the central region of the galaxy \citep[e.g.][]{Martizzi2013,vanderVlugt2019}. This process, however, has not been demonstrated to explain stellar core kinematics.   

Sinking and coalescing SMBHs in merging elliptical galaxies and their interaction with stars are so far the only process that is successful at explaining both the photometric and kinematic properties of stellar cores. During a binary merger of ellipticals, the two nuclear SMBHs sink to the centre of the remnant through dynamical friction. The surrounding stars receive energy from the coalescing SMBHs and move to higher energy orbits at larger radii or leave the merger remnant entirely. This is the dominant process for the rapid formation of stellar density cores \citep{Merritt2013,Rantala2018}. Once the two black holes form a bound binary system at the centre of the galaxy, core formation is mostly complete. The core size can be significantly increased by the repeated merger of two galaxies with pre-existing cores \citep{Rantala2019} or, possibly, the ejection of the SMBH merger remnant by a gravitational recoil kick \citep[e.g.][]{Nasim2021}. Until their final coalescence by gravitational wave emission \citep[see e.g.][for binary elliptical galaxy merger simulations with SMBHs]{Mannerkoski2019} the SMBH binaries still have 3-body interactions with stars that venture too close. This process often results in stars getting kicked out from the centre of the galaxy \citep{HillsFullerton1980}, and is termed as gravitational `slingshot' or `scouring' \citep[see e.g.][]{Merritt2013}. Throughout this paper we refer to this process as `slingshot', and consider `scouring' the overall core formation process, including the early phase dominated by dynamical friction. The slingshot process only slightly lowers the stellar density of the core. However, it works mostly on stars with radially-biased orbits coming close to the SMBH binary and, by ejecting those from the core region, it slowly develops a tangentially-biased stellar velocity distribution in the core \citep{Rantala2018}.

In this paper we investigate merger simulations presented in previous studies \citep[][]{Rantala2018,Rantala2019} with a particular focus on the slingshot process and how it affects the orbital distribution of stars. Stellar orbits are the backbone of the
structure of every elliptical galaxy, and the most massive ellipticals are, to a good approximation, gas-poor collisionless systems. Therefore, they preserve information about their formation processes encoded in their orbital structure for billions of years  \citep{Jesseit2005,Roettgers2014,Frigo2019}. In this study we demonstrate how the observed properties of the cores of massive ETGs are set in two phases: first the rapid formation of the stellar density core during the merger, and then the slower change of the orbital structure of the core towards a tangentially-biased distribution through the slingshot process, on timescales of several 100 Myr. 

In Section 2 we briefly review the simulation details, followed by the description of the orbit analysis pipeline in Section 3. The time evolution of density and stellar velocity anisotropy in our fiducial simulation is investigated in detail in Section 4. Here we demonstrate how slingshot kicks by the SMBH binary drive the core region towards negative velocity anisotropies (tangential orbits). In Section 5 we analyse the full orbital content of the simulated mergers and show which orbit classes are more likely to interact with the SMBH binary and which orbit types they become thereafter. In Section 6 we show how the orbital structure and the dynamical as well as morphological properties correlate with SMBH mass. In Section 7 we dissect our fiducial merger remnant, which has a kinematically decoupled core, and assign observable kinematic features to specific orbit classes. We also demonstrate that  the orbit analysis with a novel three-dimensional Schwarzschild modelling method \citep{Neureiter2021}, based on the kinematic maps, agrees remarkably well with our direct orbit analysis. We summarise our results and conclude in Section 8.

\section{Simulation details}
In this paper we analyse a subset of the early-type galaxy merger simulations presented in \citet{Rantala2018} and \citet{Rantala2019}.  Here we give a brief overview of the simulation code, of the initial galaxy models, and of the merger setup. The simulations were run with an extension of the {\small GADGET}3 code \citep{gadget2}, called {\small KETJU} \citep{Rantala2017}, that adds a region around every supermassive black hole where the orbits of particles are calculated at high precision and without gravitational softening, using an algorithmic regularisation technique \citep{Mikkola1999, Mikkola2008,Karl2015, Rantala2020, Wang2021}. Post-Newtonian corrections are also included, as well as relative loss terms between SMBHs. The high accuracy allows to resolve the dynamical interactions between two SMBHs and between SMBHs and the stellar population particles. The procedure also allows to adequately compute the in-spiralling of the individual SMBHs and hardening of forming SMBHs \citep[see also][for a similar integration strategy]{Karl2015} binaries as well as the interaction of the SMBHs with stars. The interactions between different stars within the high-resolution region are also calculated with the same technique, which means they do not experience gravitational softening unlike in the rest of the simulation box. This could in principle produce unrealistic interactions, but since in our case the potential of the black hole binary completely dominates the regularised region, we expect this not to impact the simulations. For a full description of {\small KETJU} we refer the reader to the detailed presentation in \citet{Rantala2017}, and to the similar approach outlined in \citet{Karl2015}.

The initial galaxy models are spherically symmetric and made of stars, dark matter, and a central black hole. The dark matter component follows a \citet{Hernquist1990} density profile, while the stellar component follows a \citet{Dehnen1993} density profile with slope $\gamma = 3/2$, which when projected reproduces approximately the commonly observed \citet{devaucouleurs} luminosity profile ($ L \propto {\rm exp}(-R^{1/4})$, see e.g. \citealp{NaabTrujillo2006} for a discussion about projected merger remnant surface density profiles). The black hole is represented by a point mass at the centre of each galaxy. 
The total stellar mass is $4.15 \times 10^{11} \, M_\odot$, while the total dark matter mass is $7.5 \times 10^{13} \, M_\odot$. The half-mass radius of the stellar component is $7 \, \rm kpc$. These parameters are chosen such that the merger remnants resemble in mass and size the observed galaxy NGC1600 \citep{Thomas2016, Rantala2018}. The dark matter fraction within the stellar half-mass-radius is always $f_{\rm DM}(r_\mathrm{1/2}) = 0.25$. The stellar particles have mass $m_\star = 1.0 \times 10^5 M_\odot$, while the dark matter particles mass $m_{\rm DM} = 7.5 \times 10^6 M_\odot$. The simulations have varying initial black hole masses from $8.5 \times 10^{8} \, M_{\odot}$ to $8.5 \times 10^{9} \, M_{\odot}$ ). The initial model is dynamically stable and has an isotropic velocity distribution with a distribution function calculated using Eddington's formula (see Chapter 4 of \citealp{BinneyTremaine}, as well as \citealp{Hilz2012,Rantala2017}).

We analyse 7 equal-mass merger simulations with varying initial SMBH masses. The two galaxy models start at a distance of 30 kpc from each other and move on nearly parabolic orbits, with a pericentre distance of $r_p \sim 0.5 \times r_{1/2}$, where $r_{1/2}$ is the half-mass radius of the initial model (7 kpc). With increasing SMBH mass we label the simulations in the following way: ETG-1-1 ($M_{\rm BH} = 1.7 \times 10^{10} \, M_{\odot}$), ETG-1-1-bh5 ($1.36 \times 10^{10} \, M_{\odot}$), ETG-1-1-bh4 ($1.02 \times 10^{10} \, M_{\odot}$), ETG-1-1-bh3 ($6.8 \times 10^{9} \, M_{\odot}$), ETG-1-1-bh2 ($3.4 \times 10^{9} \, M_{\odot}$), ETG-1-1-bh1 ($1.7 \times 10^{9} \, M_{\odot}$). The simulation ETG-1-1-nobh has no SMBHs. All details on the initial setup for these simulations can be found in \citealp{Rantala2018}, who also use the same simulation labels. 

\section{Orbit analysis}
\label{sec:orbitpipeline}
\begin{figure}
    \centering
    \includegraphics[width=\columnwidth]{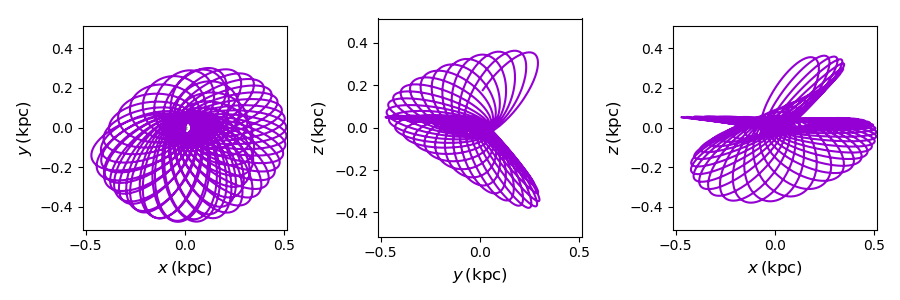}
    \includegraphics[width=\columnwidth]{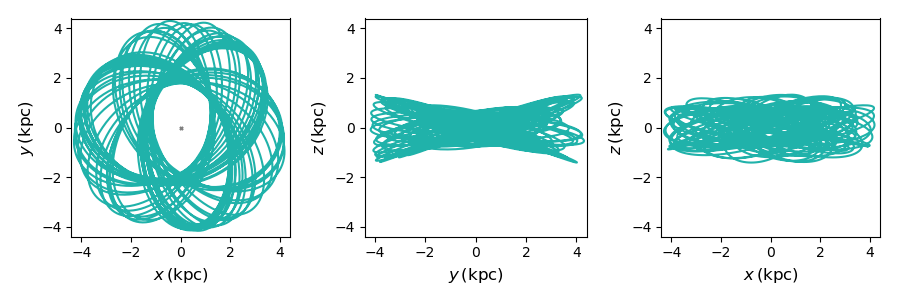}
    \includegraphics[width=\columnwidth]{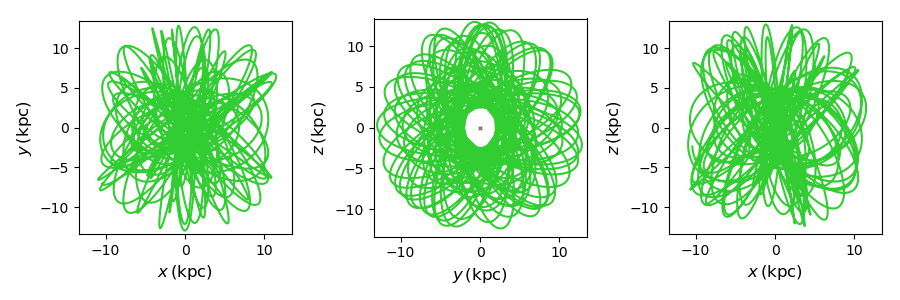}
    \includegraphics[width=\columnwidth]{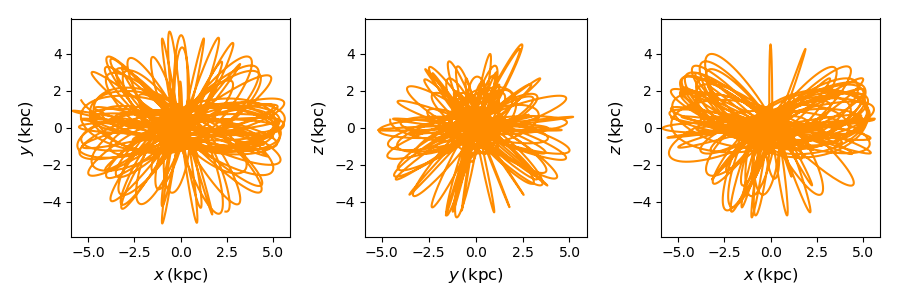}
    \caption[]{Example of four different orbit types (from top to bottom: rosette, z-tube, outer x-tube, box), each seen along the three principal axes. Tube and rosette orbits are centrophobic, in contrast to box orbits, which can interact with central SMBHs.}
    \label{fig:orbit_example}
\end{figure}

\subsection{Orbit types}\label{sec:orbittypes}
In this paper we classify the stellar orbit types of all stellar particles in the simulations. These include all orbits generally available in triaxial potentials \citep{BinneyTremaine}, with the addition of rosette orbits near the SMBHs \citep[see e.g.][]{Neureiter2021}. We specifically distinguish: 
\begin{itemize}
\item z-tube or short-axis tube orbits - these orbits rotate around the minor (z) axis of the galaxy, and are the typical rotational orbits common in disc galaxies.
\item x-tube or long-axis tube orbits - orbits that rotate around the major (x) axis of the galaxy. They are common in prolate elliptical galaxies, which sometimes display net major-axis rotation in their projected kinematic maps \citep[e.g][]{Krajnovic2018}. Prolate galaxies are also known to be regularly produced in mergers of gas-poor galaxies \citep[see][for idealised and cosmological simulations]{Barnes1996,Naab2003,Jesseit2005,Cox2006,Jesseit2007,Hoffman2010,Ebrova2015,RodriguezGomez2015,Li2018,Schulze2018}. X-tube orbits further subdivide into two categories depending on their shape: inner and outer x-tubes. Inner x-tubes have a concave shape, meaning that stars on these orbits move almost radially when far from the centre. Outer x-tubes on the other hand are convex-shaped, and are analogous to z-tube orbits, but are rather rotating around the x-axis.
\item rosette or spherical orbits -  typical orbits of a spherically symmetric potential dominated by a point mass in the centre. They resemble Keplerian orbits, and have a stable orbital plane, which can be oriented in any direction (if the potential is spherically symmetric, each component of the angular momentum is conserved). They are common near SMBHs at the centre of galaxies even if the entire system is triaxal. 
\item \pibox\ orbits -  orbits with no net angular momentum. Stars on these orbits typically move radially in the galactic potential and can get arbitrarily close to the centre of the potential. Their motion along the three principal axes is not resonant (hence the $\pi$ in their name).
\item boxlet orbits - box orbits (no net angular momentum) that show a resonance between their frequencies along the three main axes. Low level resonances give rise to familiar shapes, such as fish orbits (3:2). These orbits move radially, but can avoid the centre.
\item irregular orbits - chaotic orbits that are not bound in phase space by integrals of motion.
\item not classified -  orbits that our classification scheme fails to classify. More on this later in this section.
\end{itemize}
In a triaxial potential certain orbit classes, such as intermediate axis/y-tubes, are not stable (but see \citet{Neureiter2021}). Several papers \citep{Merritt1999, Poon2001, Merritt2011, Merritt2013} introduced new orbit families for triaxial potentials with a SMBH at the centre, in particular pyramid orbits. Similarly to box orbits, pyramid orbits have different frequencies along the different axes. The main difference is that when they pass by the centre of the potential their orbit around the black hole makes a 180 degrees turn back, while a normal box orbit would have continued onwards. Because of this they form a pyramid-like shape. In our classification scheme these orbits would be classified as boxes, and like box orbits, stars on pyramid orbits get arbitrarily close to the centre of the potential given enough time. Because of this they are prime subjects for interactions with a SMBH binary. 

\subsection{Orbit classification}
Many studies of stellar orbits of simulated galaxies \citep[e.g.][]{Barnes1996,Hoffman2010} have used a simple and efficient orbit classification scheme introduced by \citet{Barnes1992}.  
In this study we use instead a classification procedure presented in \citet{Jesseit2005}. This procedure is based on the spectral scheme of \citet{Carpintero1998}, which has been applied to idealised and cosmological simulations \citep[e.g.][]{Naab2006,Thomas2009,Bryan2012,Roettgers2014,LiB2015,Frigo2019} 
\subsubsection{Representation of the potential and orbit integration}
We classify stellar particle orbits  following \citet{Carpintero1998} with a time integration in simulated galaxy potentials as implemented in \citet{Jesseit2005,Naab2006}, \citet{Roettgers2014} and \citet{Frigo2019}.
We reconstruct the potential generated from the simulation particle data with  with   self-consistent field (SCF) method \citep{Hernquist1992} using the \citet{Hernquist1990} profile as a basis function at zeroth order: 
\begin{equation}
\rho_{000} = \frac{M}{2 \, \pi \, a^3} \frac{1}{\frac{r}{a} (1 + \frac{r}{a})^3}
\end{equation}
\begin{equation}
\Phi_{000} = - \frac{G \, M}{r + a},
\end{equation}
where $a$ is the scale parameter and M is the total mass. Higher-order terms determine the detailed radial and angular shape of the potential \citep{Hernquist1992}. In our analysis we limited the maximum radial and angular order of this expansion to $n_{\rm max} = 18$ and $l_{\rm max} = 7$ respectively. 
Before the SCF fit, the snapshot is centred on the centre of mass of the black hole binary (or on the point with the highest stellar density if there are no black holes), and oriented according to the reduced inertia tensor (see \citealp{Bailin2005}) of the stars, so that the galaxies' long axis is aligned with the $x$ axis and the short axis is aligned with the $z$ axis. The potential fit is applied to the simulation snapshot {\it without including the SMBH binary}, since its point-mass  potential at the centre cannot be fitted with our basis function set. The potential of the binary is instead added for the orbit integration as a point mass potential. 

The orbit of each stellar particle is then integrated with an eighth -order Runge-Kutta integrator for 50 orbital periods in the frozen analytic potential starting from the position and velocity in the respective snapshot. The integration is long enough to identify the orbital family of the particle, but short enough that quasi-regular orbits (see Chapter 4 of \citealp{BinneyTremaine}) do not diverge from regular phase-space regions \citep[see][for a discussion]{Roettgers2014}. The orbits are integrated in a static, analytical potential without 2- and 3-body interactions in particular without the SMBH binary). Stellar particles that pass close to the SMBH binary in the simulation might get kicked out of the galaxy. In these cases the orbit classification pipeline can tell us which orbit family the particle would have belonged to if there was a single black hole instead of the black hole binary. We investigate this in Section \ref{sec:scouring}.

\subsubsection{Fourier analysis and classification}
For the orbit classification, the frequencies of each particle's motion along the three axes are calculated through a Fourier transform (using the {\small FFTW} algorithm of \citealp{fftw}). The frequencies are then checked for resonances \citep{Carpintero1998}. A 1:1 resonance indicates a tube orbit, z-tube or x-tube, depending on the axes which display resonance. In the case of an x-tube, the convexity over the entire orbit is calculated here, in order to determine  whether it is an inner x-tube (concave) or an outer x-tube (convex). If the orbit shows a 1:1:1 resonance along all axes, it is classified as a planar rosette orbit, typical of spherical potentials. If the frequencies are in a $m$:$n$ ratio, the orbit is classified as a boxlet (resonant box orbit). If the ratios between the frequencies are not simple integers (the code checks $m < 29, n < 10$, if $m:n$ is the resonance), the orbit is classified as a \pibox, where $\pi$ denotes the `irrational' ratio.  If there are more than three base frequencies it is classified as irregular. Finally, the classification of the full orbit is compared with the one of two partial segments of it. If they do not match with each other, the orbit is considered `not classified'. If the orbit has a 1:1 resonance between the $x$ and $z$ axes (a `y-tube') it is also categorized as not classified, since these orbits cannot be stable in a triaxial potential. It has however been shown, using the ETG-1-1 simulation studied in this paper, that such intermediate-axis tubes can exist and be stable for a long time (up to a Gyr), even though they are rare and eventually become chaotic \citep{Neureiter2021}. In addition, we also track the pericentre, apocentre, and mean radius of each orbit. 

\section{Secular evolution of a galaxy core by SMBH binary slingshots} \label{sec:scouring}

\begin{figure}
    \centering
    \includegraphics[width=\columnwidth]{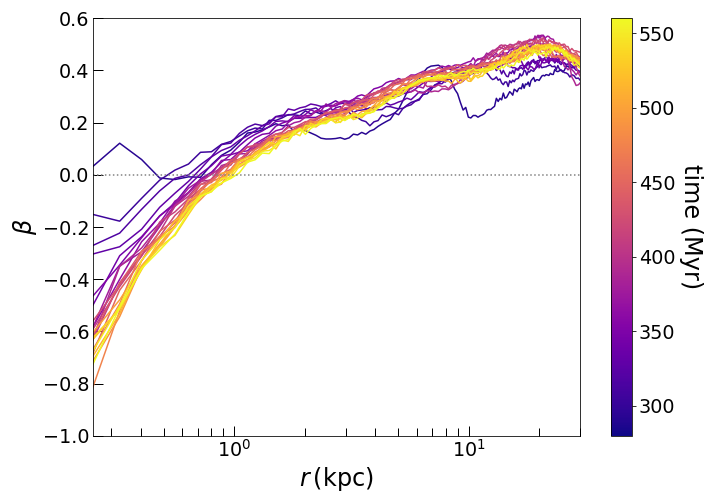}
    \caption[]{Stellar velocity dispersion anisotropy profile of ETG-1-1 at every snapshot between 250 and 550 Myr. The different colors from blue to yellow indicate different times. With increasing time, the profile evolves increasingly more towards tangential anisotropy (i.e. negative $\beta$) within the central kpc.}
    \label{fig:anisotime2}
\end{figure}

\begin{figure}
    \centering
    \includegraphics[width=\columnwidth]{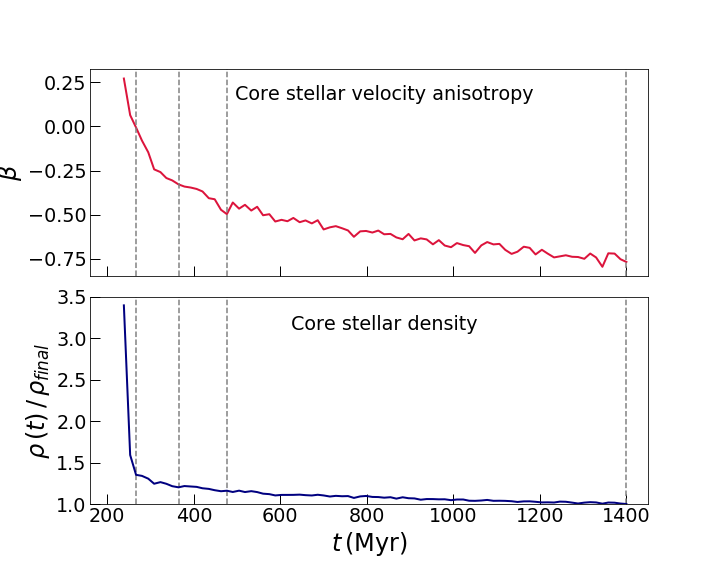}
    \caption[]{Top panel: time evolution of the central stellar velocity dispersion anisotropy within $r_{\mathrm{SOI}}  = 0.86 \, \rm kpc$ (the core size at $t = 300$ Myr)  as a function of time. Bottom panel: time evoluton of the stellar density inside the same region as a function of time. While the density core is formed immediately after the merger of the central regions of the galaxies \citep[see][]{Rantala2018} at $ t \sim 250$ Myr, the anisotropy keeps decreasing over the next Gyr of evolution driven by stellar slingshots with the SMBH binary. The vertical dashed lines indicate the times at which the in-depth orbit analysis is performed (see Section \ref{sec:orbitresults}).}
    \label{fig:anisorho}
\end{figure}

\begin{figure}
    \centering
    \includegraphics[width=\columnwidth]{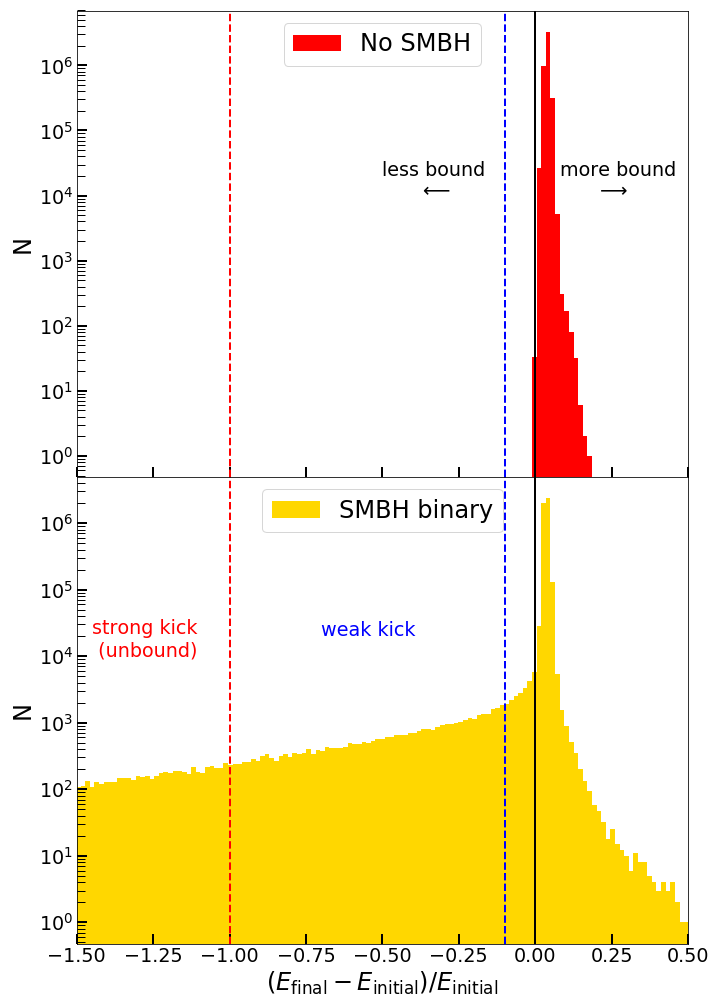}
    \caption[]{Histograms of the relative change in energy between 300 and 600 Myr of every stellar particle inside the stellar half-mass radius for the simulation without SMBHs  (ETG-1-1-nobh, top panel) and the simulation with the most massive SMBHs (ETG-1-1, bottom panel). Bound particles have negative total energy and particles with positive $\Delta E / E$ have therefore become more bound. In the simulation without SMBHs the binding energies of the stars change very little. In the SMBH case about 2 per cent of the stars gain energy from interactions with the SMBH binary. We separate those particles into `strong kicks' ($\Delta E / E < -1$) after which stars become unbound, and `weak kicks' ($\Delta E / E < -0.1$) for stars that receive an energy increase of more than 10\% but not strong enough to become unbound. }
    \label{fig:energychange}
\end{figure}

\begin{figure}
    \centering
    \includegraphics[width=\columnwidth]{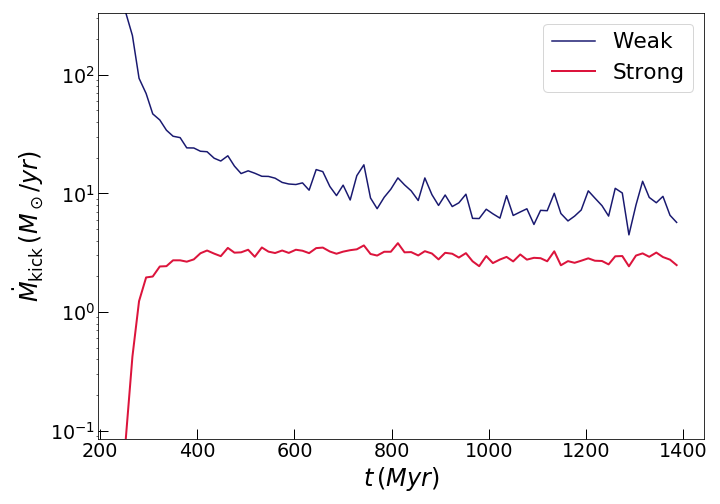}
    \includegraphics[width=\columnwidth]{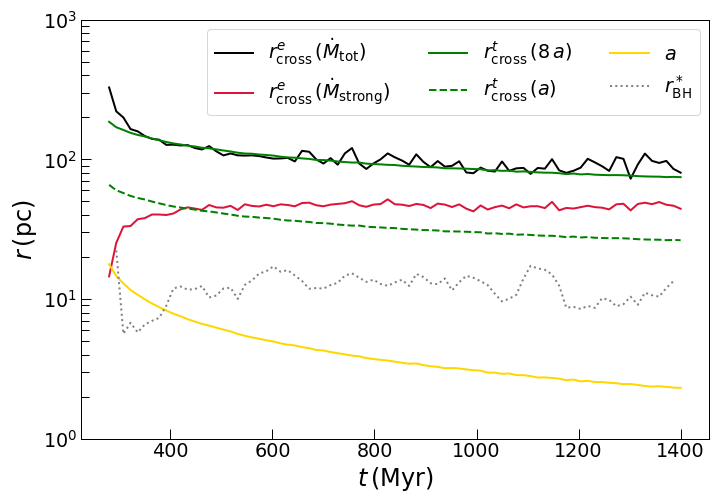}
    \caption[]{Top panel: Mass rate of kicked stars in $M_\odot \rm yr^{-1}$ as a function of time, calculated with a time resolution of 14 Myr. Weak and strong kicks are distinguished in blue and red, respectively. The strong kick rate quickly approaches a constant value and stars with strong kicks leave the galaxy at a rate of $2 M_\odot \rm yr^{-1}$. Many weakly kicked stars experience multiple kicks, often resulting in a final strong kick. Bottom panel: effective cross section radius \rcross\ for weak and strong kicks (see Eq. \ref{eq:rcrosse}), compared with the theoretical expectation from Eq. \ref{eq:rcrosst} (green), the semi-major axis of the black hole binary (yellow) and the average displacement between the binary and the stellar centre of mass \rbhstar\ (dotted grey). }
    \label{fig:kickrate}
\end{figure}

\begin{figure}
    \centering
    \includegraphics[width=\columnwidth]{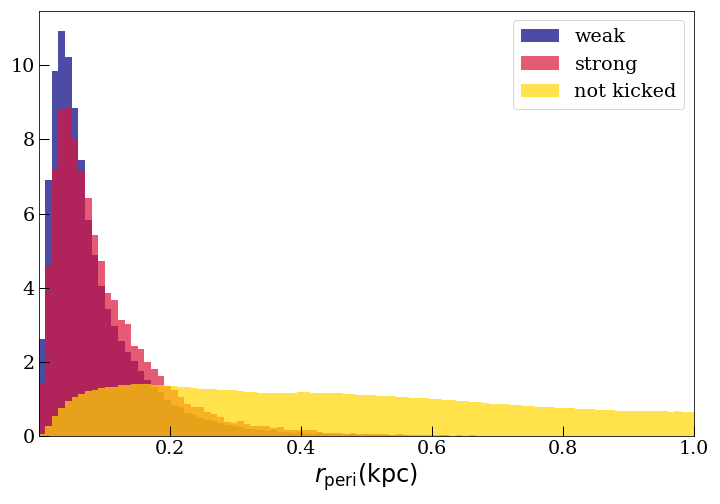}
    \caption[]{Histogram of the pericentre distance of stellar particles at t=300 Myr, distinguishing particles that by the last snapshot at t=1400 Myr will receive a weak kick (blue), a strong kick (red) and will not get kicked (yellow). The histograms are all normalized to one. Particles that will get kicked predominantly have small initial pericentre distances $(< 0.2 \ \rm kpc)$, and there is no difference in pericentre distance between weak and strong kicks.}
    \label{fig:perikicked}
\end{figure}

\begin{figure}
    \centering
    \includegraphics[width=\columnwidth]{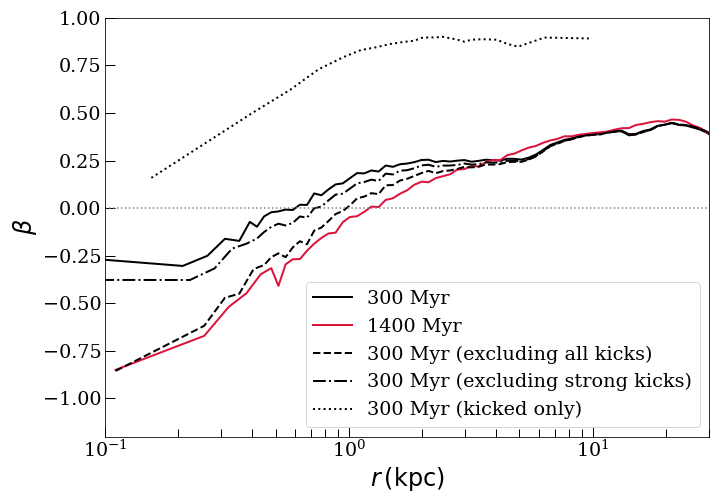}
    \caption[]{Anisotropy profile of ETG-1-1 at 300 Myr (right after the merger of the galaxy centres, blue solid line) and at 1400 Myr (red solid line). The blue dashed line shows the anisotropy profile at  $t = 300$ Myr excluding all particles receiving weak or strong kicks in the future (by 1400 Myr). The good agreement with the real final profile indicates that the resulting anisotropy can be entirely accounted for by removing stars having interacted with the SMBH binary.Initially,  the kicked particles (black dotted line) have radially biased orbits and just leave the galaxy centre after the kicks. Strongly kicked particles alone cannot account for the final anisotropy profile (blue dotted line).}.
    \label{fig:anisotime}
\end{figure}

\subsection{The two phases of core formation}
A galactic major merger without gas reshapes the mass and velocity distribution of the progenitor galaxies through violent relaxation \citep[see e.g.][and references therein]{Hilz2012, Hilz2013}, and after a few crossing times the system settles into a new equilibrium. However, interactions between stars and a SMBH binary in the centre of the galaxy (slingshots) result in a slow change in the core properties - a process which can last for billions of years.

In Fig. \ref{fig:anisotime2} we plot the stellar velocity anisotropy $\beta$ as a function of radius, for different times (increasing from blue to yellow) after the merger.
The anisotropy $\beta$ is defined as:
\begin{equation}\label{eq:aniso}
\beta \, (r)=1-\frac{\sigma_\phi^2 + \sigma_\theta^2}{2 \, \sigma_r^2},
\end{equation}
where $\sigma_\phi$, $\sigma_\theta$ and $\sigma_r$ are the velocity dispersion along the $\phi$, $\theta$ and radial directions. $\beta$ is positive when orbits are biased towards being radial, and it is negative when orbits are biased towards being tangential. The first profile (blue) is shown at $t \sim 260$ Myr when the galaxy centres have merged and the SMBHs have formed a hard binary. At this point $\beta$ is close to zero in the core region ($ r \lesssim 1$ kpc) and positive (radial bias) further out. The wave patterns at larger radii are produced by shells of material pushed out during the coalescence of the galaxy centres. With increasing time, $\beta$ decreases to more negative values (tangential bias) in the core. In \citealp{Rantala2018} it was shown that $\beta$ remained constant after artificially merging the SMBH binary a bit after $\sim 300$ Myr. This is direct evidence that the existence of a SMBH binary and its interaction with stars are driving the velocity dispersion evolution.

It is interesting to note that this process is {\it slow} compared to the formation of the low-density core which has almost fully formed immediately after the central merger is complete.
In Fig. \ref{fig:anisorho} we compare the time evolution of the central stellar velocity anisotropy to the central stellar mass density, both calculated in a sphere of radius $r_{\rm SOI}(t = 300 \, {\rm Myr})=0.86 \, \rm kpc$, the initial black hole sphere of influence. Here $r_{\rm SOI}$ is defined as the radius inside which the amount of stellar mass equals the mass of the SMBH binary.
The central velocity anisotropy drops to negative values very soon after the merger ($\beta \sim -0.3$ at $t =400$ Myr) and continues to decrease for another Gyr, albeit more slowly, reaching $\beta = -0.75$. In contrast, the central density drops rapidly after the central merger and only decreases by $ \sim 15$ per cent during the next Gyr of evolution.
This suggests that the central flattening in the density profile and the anisotropy evolution towards a tangential velocity distribution have two different timescales, even though both are connected to the presence of SMBH binaries. The low-density core is produced rapidly when the sinking SMBHs become bound, even before they form a hard binary system at $t \sim 260$ Myr \citep[e.g.][]{Rantala2018}. The velocity anisotropy, however, continues to decrease after the SMBH binary has become hard.  After that, the black hole binary slowly pushes stars on radial orbits away from the core through slingshot interactions. This effect is negligible in the density profile, but prominent in the anisotropy due to its definition (i.e. with the radial velocity dispersion in the denominator, see Eq. \ref{eq:aniso}).

\subsection{Slingshot kicks on single particles}\label{sec:kick_energy}
To better understand the slingshot process, we now study how the SMBH binary affects single stellar particles in the simulation. In Fig. \ref{fig:energychange} we plot a histogram of $\Delta E =  (E_\mathrm{600} -E_\mathrm{300})/ E_\mathrm{300}$, the change in total energy (potential plus kinetic) between t = 300 Myr and t = 600 Myr, relative to the initial energy, for every particle inside the half-mass radius at t=300 Myr. Here all particles at t =300 Myr have a negative total energy and are thus bound. Particles with a negative energy change have gained energy and have become unbound if $\Delta E < -1$.  We plot the $\Delta E$ distributions both for the simulation without black holes (ETG-1-1-nobh, top panel) and the simulation with the most massive black holes (ETG-1-1, bottom panel). In the case without SMBHs very little energy change occurs. The stars become slightly more bound (positive $\Delta E / E$) on average \citep[see e.g.][]{Hilz2012}. In the case with SMBHs, a few star particles are more bound to the SMBHs but there is a significant tail of particles with a negative relative change, meaning that they become less bound by interacting with the SMBH binary. We separate these particles in two groups:
\begin{itemize}
\item $\Delta E / E < -1$, or `strong kicks': these particles become unbound after one or more interactions with the SMBH binary.
\item $-1 < \Delta E / E < -0.1$, or `weak kicks': these particles remain bound after interacting with the SMBHs, but receive a significant boost in energy (more than 10\%), which pushes them to larger radii and affects their orbital type.
\end{itemize}
Strong and weak kicks happen to only 0.8\% and 1.2\% of simulation stellar particles respectively, but since most of these particles were situated in the central regions before the kick, they have a strong effect on the central dynamical structure. 
Many particles experience a relative energy change smaller than 10\%, but we do not consider them kicked, since this small change would not affect the overall kinematics.

In the top panel of Fig. \ref{fig:kickrate} we plot the mass rate of kicked stars as a function of time, distinguishing weak and strong kicks in blue and red, respectively. The rates are calculated comparing subsequent snapshots with a time resolution of 14 Myr (the difference in time between two snapshots). Weak kicks are very common in the final phases of the galactic merger and the rate levels out to an almost constant value of about 10 $M_\odot /  \rm yr$. Instead, strong kicks are rare in the beginning but with the hardening of the SMBH binary the rate rapidly increases to a constant value of about 3 $M_\odot / \rm yr$. This is caused by the relatively slow velocity of the black holes at the beginning of the simulation ($\sim 500$ km/s) compared to the escape velocity from the centre of the galaxy ($\sim 2500$ km/s). Once the binary shrinks and the black hole velocity reaches values above 1000 km/s strong kicks become more common.
The different kicked mass rates for weak and strong kicks can be represented in terms of an effective cross section. Given a kick rate $\dot{M}$, an average stellar core density $\rho_{\rm c}$ and an average core stellar velocity dispersion $\sigma_{\rm c}$, we can define the effective cross section as:
\begin{equation} \label{eq:rcrosse}
    \Sigma_e (\dot{M}) = \frac{\dot{M}}{  \rho_{\rm c} \, \sigma_{\rm c}} \, .
\end{equation}
This value represents the expected cross section for interactions that result in a slingshot kick. Assuming that this cross section is circular we can calculate its radius as $r_{\rm cross}^e = \sqrt{\Sigma_e/\pi}$. In the bottom panel of Fig. \ref{fig:kickrate} we plot $r_{\rm cross}^e$ for all kicks (black) and for strong kicks only (red). The cross section radius for general kicks is about 100 pc wide, gradually decreasing due to the shrinking of the BH binary. This is similar to the value for weak kicks, since they outnumber the strong kicks. The cross section radius for strong kicks, $r_{\rm cross}^e (\dot{M}_{\rm strong})$ is lower, at about 50 pc. Similarly the kicked mass rate it starts low and increases quickly to a constant value. These lines are compared with a theoretical prediction for the three-body scattering cross section \citep{Celoria2018}:
\begin{equation} \label{eq:rcrosst}
    \Sigma_t (p_{\rm max}) \simeq 2 \, \pi  \frac{G \, M_{\rm binary} }{ \sigma_c^2} p_{\rm max}  \, ,
\end{equation}
where $p_{\rm max}$ is the largest pericentre radius for an interaction to be considered in the cross section. In the bottom panel of Figure \ref{fig:kickrate} we show $r_{\rm cross}^t \, (a) = \sqrt{\Sigma_t \, (a) / \pi }$ (dashed green line) and $r_{\rm cross}^t \, (8 \, a)$ (solid green line), where $a$ is the binary semi-major axis (shown in yellow). $r_{\rm cross}^t \, (8 \, a)$ has similar values to the total effective cross section $r_{\rm cross}^e (\dot{M}_{\rm tot})$. This suggests that stellar particles are likely to receive a weak kick
from the SMBH binary when they pass by $\sim 8$ times its semi-major axis, while they need to reach within $1 - 2 \, a$ to receive a strong kick. However, the movement of the binary within the stellar core might also increase the number of kicks. The many interactions between the binary and surrounding stars cause a Brownian-like motion of the binary over time, which allows more stars to get kicked \citep{Chatterjee2003}. The dotted grey line shows \rbhstar, the moving average of the displacement between the SMBH binary and the centre of mass of the stellar component. Throughout the whole simulation \rbhstar\ oscillates around 10 pc, which is larger than the binary semi-major axis. This probably contributes to making the strong kick rate constant, rather than declining, together with the increasing speed of the black holes as the binary semi-major axis shrinks down.  

In Fig. \ref{fig:perikicked} we show a histogram of the pericentre radii of particles in the first snapshot (t=300 Myr), distinguishing particles that will receive a weak kick (blue), a strong kick (red), or that will not be kicked at all (yellow) by the final snapshot of the simulation (t=1400 Myr). These values are taken from the analytically-integrated orbit (see Section \ref{sec:orbittypes}) in the static potential of the first snapshot, so it might not correspond to the actual closest approach between the star particle and the binary in the simulation, but they nevertheless allow us to characterize the orbit of each particle. The histograms are normalized so that they have the same area.  Kicked particles tend to have small pericentre radii ($\rperi < 0.2$ kpc), which allows them to have close encounters with the SMBH binary. There is no difference in \rperi\ between future weak and strong kicks, even though we would expect smaller \rperi\ values for strong kicks. This is likely due to the movement of the SMBH binary within the stellar core, which is not accounted for in these analytically-integrated orbits.
Nevertheless, the strength of the interaction does not depend only on \rperi. 
Another factor is the alignment between a star's orbit and the orbit of the SMBH binary. If the orbit is co-rotating with the binary, during the closest approach the star has a longer time to interact with the black holes, and is thus more likely to get kicked. 
We measured that co-rotating stellar particles are 9\% more likely to get kicked, although the percentage increases when considering only particles whose orbit is on the same plane as the binary.

\subsection{Slingshot kicks and velocity anisotropy}

We start with the hypothesis that the SMBH binary can affect the central velocity anisotropy $\beta$ by removing the kicked stars from the centre. To test this we show in Fig. \ref{fig:anisotime} the anisotropy profiles right after the merger (300 Myr, solid black line) and at the end of the simulation (1400 Myr, solid red line). We also show the $\beta$ profile of the first snapshot (300 Myr) without all particles which will receive a kick (as defined in Section \ref{sec:kick_energy}) by the SMBH binary between 300 and 1400 Myr. The resulting profile (dashed line) shows a negative central anisotropy and a profile which is very similar to the real profile at 1400 Myr (red line). This shows that the removal of particles which have interacted with the SMBH binary is the dominant process for lowering the central anisotropy. Here all kicked particles are relevant. If we only excluded strongly kicked particles the change in $\beta$ is not strong enough (dash-dotted line). With the dotted line we also show the anisotropy profile of the particles that will receive a kick, excluding the rest of the galaxy. They show high $\beta$ values typical of radially-biased orbits. While many of the kicked stars do not leave the galaxy entirely, they move to larger radii where their relative contribution to the overall anisotropy profile is smaller, because of the larger number of particles.

\section{The orbit connection}\label{sec:orbitresults}

\begin{figure}
    \centering
    \includegraphics[width=\columnwidth]{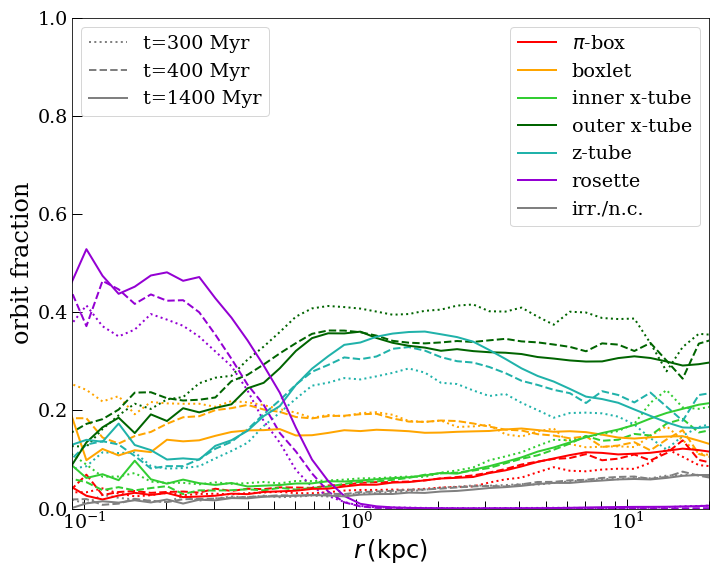}
    \caption[]{Fraction of the different orbit types as a function of radius, for simulation ETG-1-1, at times 300 Myr (dotted), 400 Myr (dashed) and 1400 Myr (solid). The central region, where the SMBH binary resides, is dominated by centrophobic tube-like orbits (rosette, x-tubes, z-tubes). With time the rosette orbit fraction slightly increases whereas the boxlet fraction shows the strongest decrease.}
    \label{fig:timeradialfreq}
\end{figure}

\begin{figure}
    \centering
    \includegraphics[width=\columnwidth]{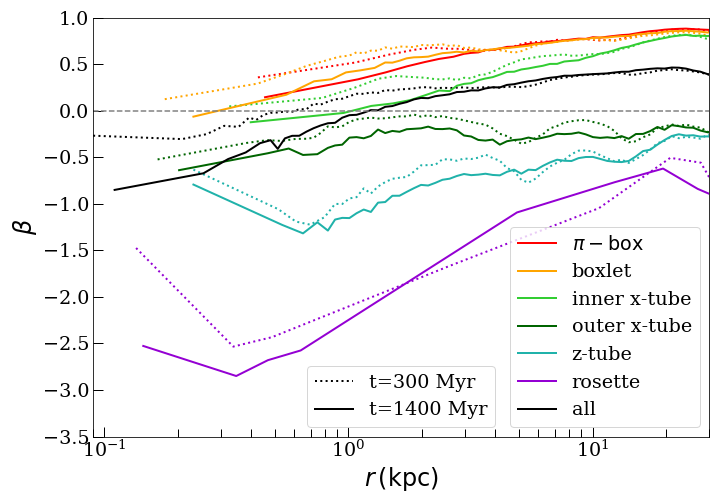}
    \caption[]{Anisotropy profile for stellar particles of different orbit classes for ETG-1-1 (coloured lines), and of the whole galaxy (black line). More tangentially-biased orbit types (rosette, z-tubes, outer x-tubes) have strongly negative values of $\beta$, while radially-biased orbit  types ($\pi$-boxes, boxlets, inner x-tubes) have positive values. }
    \label{fig:aniclass}
\end{figure}

In the previous Section we have established that stellar kicks due to interactions with the SMBH binary drive the secular change in the orbital anisotropy in the core of the galaxy. While anisotropy is a useful simple parameter to describe the dynamical state of the core, we now provide the full picture by analysing the underlying orbital composition of the simulated merger remnants. We use the orbit analysis pipeline described in Section \ref{sec:orbitpipeline} to get the orbital structure at four different snapshots: $t =$ 300, 400, 500 and 1400 Myrs of the ETG-1-1 simulation. These snapshots are marked in Fig. \ref{fig:anisorho} with the vertical dashed lines.

\subsection{Stellar velocity anisotropy and orbits}

In Fig. \ref{fig:timeradialfreq} we show the fractions of different orbit families at different radii, and how they change over time. The different linestyles represent different snapshots (300 Myr for dotted, 400 Myr for dashed, and 1400 Myr for solid), while the different colors represent different orbit families. In general the galaxy is dominated by x-tube orbits (light and dark green) in the outer parts, which account to 40-50\% of all orbits beyond 1 kpc from the centre. Most of these orbits are outer x-tubes (dark green), but there is also a non-negligible fraction of inner x-tubes (light green), especially beyond the effective radius of the galaxy. Z-tube orbits (blue) are common outside of the core, with a fraction of about 30\%. The boxlet (resonant box, orange) is almost constant with radius $\sim$ 20\%. \pibox\ (non-resonant box, red) orbits are rare, accounting for about 5\% of all orbits, and declining towards the core. The core is dominated by rosette orbits (violet, 40-50\%), as one would expect inside the SMBH sphere of influence, which varies from 0.86 kpc at t=300 Myr to 1.01 kpc at 1400 Myr. Irregular orbits are rare, accounting for less than 5\% of all orbits. 

The orbit fractions change with time. Between 300 and 400 Myr the fraction of outer x-tube orbits in the outskirts decreases by almost 10\%, while z-tubes slightly increase; this is because at 300 Myr the galaxy outskirts are not yet fully settled after the merger. The other changes are caused by the interactions between stars and the SMBH binary. The fraction of boxlet orbits drops in the core, as they are more likely to gain energy from the SMBH binary and move to larger radii or leave the galaxy entirely. Tube and rosette orbits instead increase in percentage as they are less likely to interact with the SMBHs. The fraction of \pibox\ orbits is low from the beginning but stays almost constant. 

The changes in the fraction of different orbital classes are connected to the changes in the anisotropy profile shown in Fig. \ref{fig:anisotime}. We plot the same  anisotropy profile of ETG-1-1 at t=1400 Myr and t=300 Myr in Fig. \ref{fig:aniclass} (black lines) and now split the profiles into the different orbit classes that we have defined in Sec. \ref{sec:orbittypes}. As expected, rosette orbits (pink) are highly tangentially biased, with anisotropy values typically below $\beta < -1$. Z-tube orbits (blue) are also tangentially biased at all radii. X-tubes (green), however, have a different behaviour depending on their subclass. Outer x-tubes are similar to z-tubes except for their orientation, and are therefore tangentially biased. Inner x-tubes instead move radially far from the inner region, and are therefore slightly radially biased.  \pibox\ and boxlet orbits (red and orange, respectively) are always radially-biased, and have very similar $\beta$ values to each other. The global anisotropy profile (black lines) results from a superposition of these orbital components, weighted by their mass contribution at each radius.
The anisotropy profiles for each orbit class change only moderately between 300 Myr and 1400 Myr. At 1400 Myr the $\beta$ values for boxlet and \pibox\ orbits are a bit closer to zero, suggesting that even among box orbits the most radially-biased orbits are kicked. This effect, together with the decrease in the fraction of central boxlet orbits seen in Fig. \ref{fig:timeradialfreq}, causes the shift in anisotropy towards more tangential (negative) values.

\begin{figure*}
    \centering
    \includegraphics[width=\textwidth]{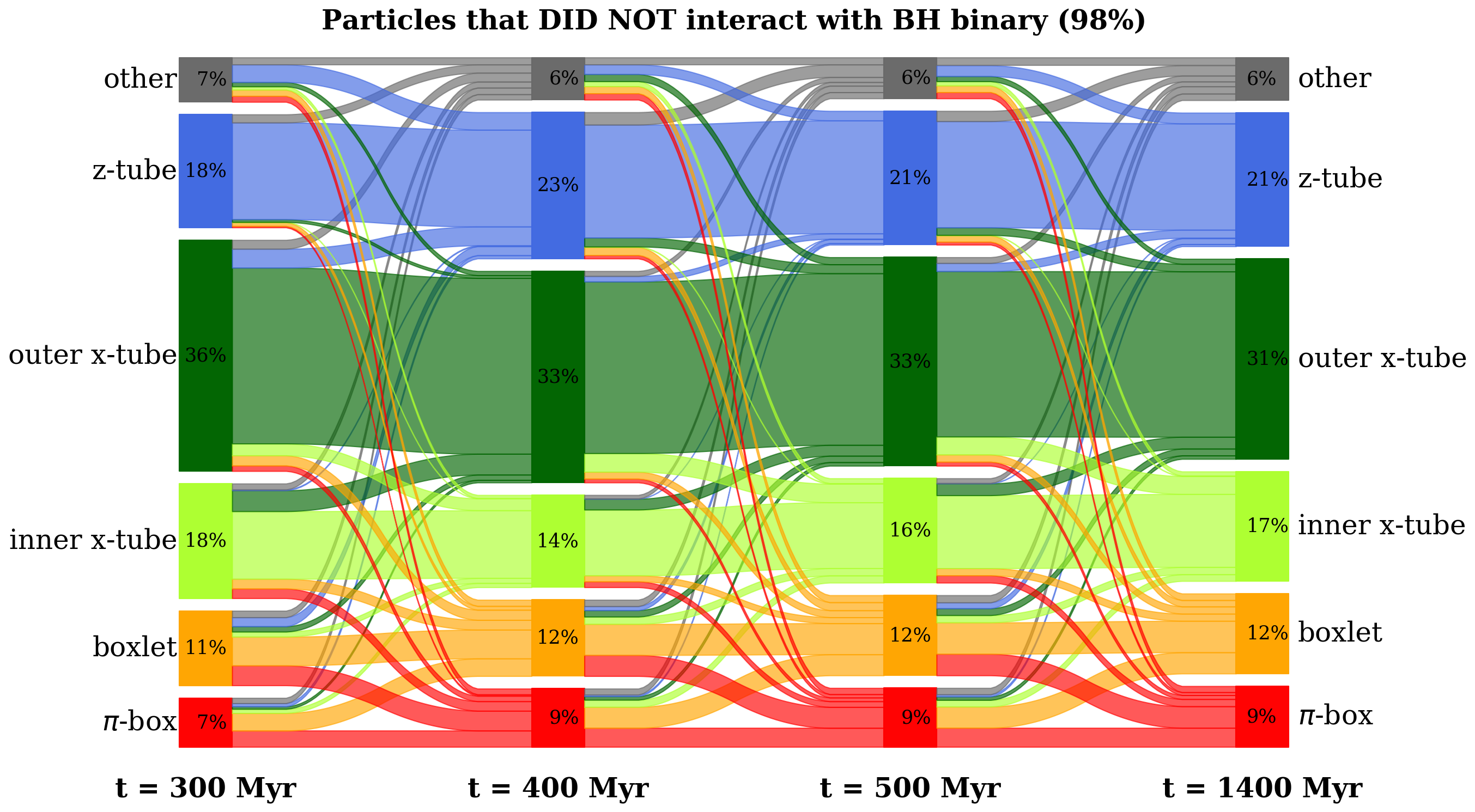}
    \includegraphics[width=\textwidth]{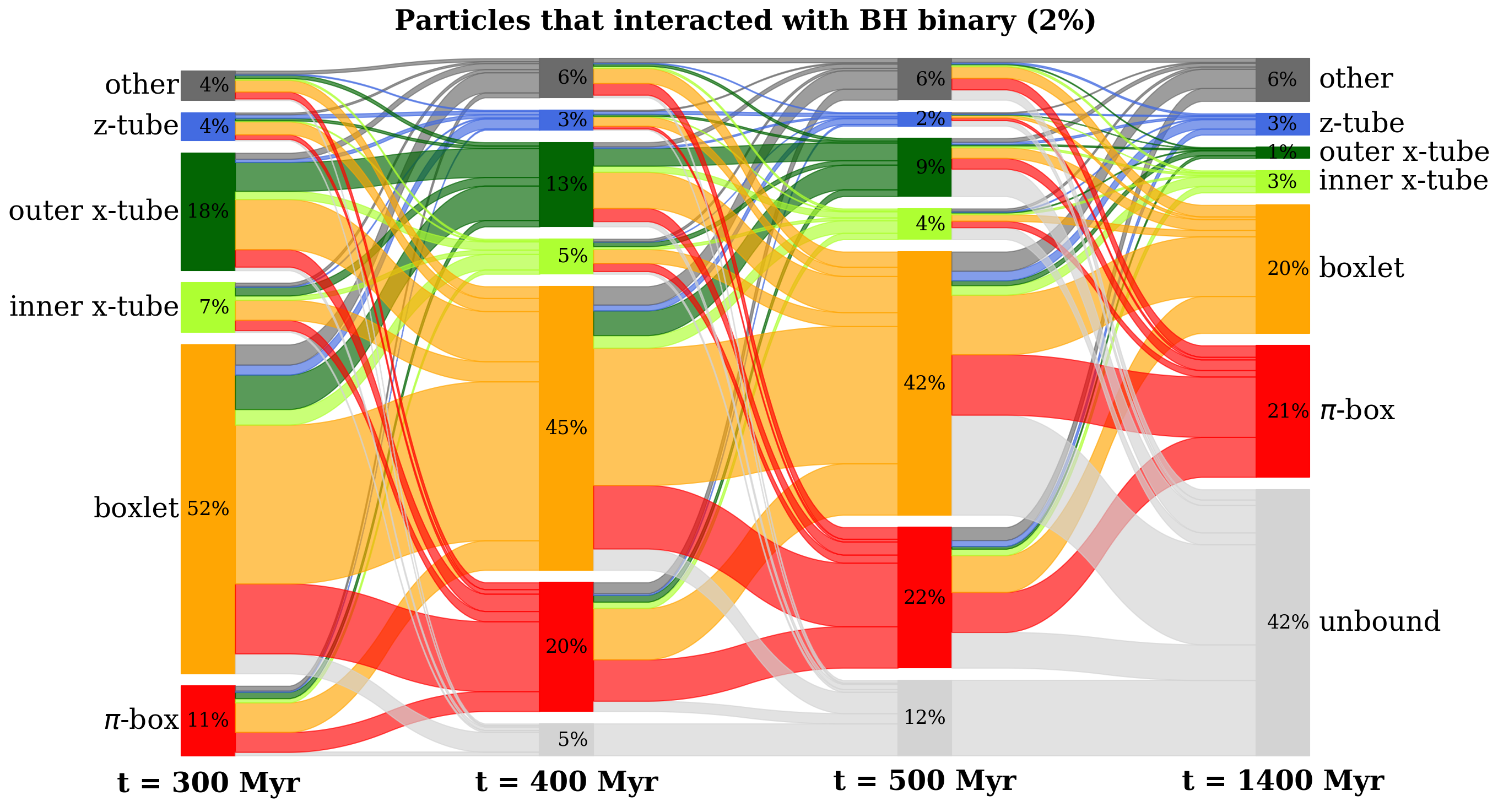}
    \caption{From left to right we show the time evolution of the fraction of orbit classes of particles that have not interacted with the SMBH binary (top panel) and that have interacted with it (bottom panel) at any point during the simulation at four different times (300, 400, 500 and 1400 Myr, dark histogram bars). We also indicate the flow between different orbit classes between the respective snapshots.
    The orbital composition for non-interacting particles stays roughly constant, while interacting particles see a significant change: 41\% become unbound, while the remaining ones are more likely to become box orbits, especially \pibox\ ones. Particles becoming unbound (grey) are predominantly taken from the box orbit reservoir. Tube orbits are unlikely to directly become unbound, but often first shift to box-like orbit classes, and can then become unbound in a second kick. }
    \label{fig:sankeykicked}
\end{figure*}

\subsection{The SMBH binary driven change of central orbits}\label{sec:kick_sankey}
Based on the results in the previous sections we expect stellar particles on orbits with typically low angular momentum (\pibox\ and boxlets) to predominantly interact with the SMBH binary. The corresponding analysis is shown in Fig. \ref{fig:sankeykicked}. The top panel shows particles that do not interact with the SMBH binary, while the bottom panel only shows the particles that have interacted at any time between 300 and 1400 Myr (both weak and strong kicks, as defined in Fig. \ref{fig:energychange}). In each panel the darker histogram columns indicate the fractions of different orbit families in the four snapshots we considered (300, 400, 500 and 1400 Myr). The lighter coloured areas show how these orbit fractions change between the different snapshots. The less common orbit families have been grouped together into `other' (dark grey)
in order to increase the readability of the plot. Among particles that do not interact with the SMBHs (top panel) there is not much change in orbit type. There is a small decrease in the fraction of x-tubes between the first and second snapshots (from 55\% to 48\%), which corresponds to an increase in the fraction of z-tubes. This is due to the outer parts of the galaxy not being completely settled by $t = 300 \, \rm Myr$. Other changes in orbit type are mainly caused by the assumptions inherent in our orbit analysis, mainly the assumption of a static potential, which we know is not accurate (because the central stellar density decreases with time and the BH binary is moving relative to the stellar core).
Among particles that do interact with the SMBH binary the orbit type is much more unstable. Here we defined as `unbound' particles that have received strong kicks and have left the galaxy (shown in grey). At all times interacting particles come mainly from box orbit types (boxlet and \pibox), with an initial value of 64\% even though this is not the dominant orbit family in the system overall (only 21\% of all orbits are boxes). Instead, only 32\% of the interacting orbits are x-tubes, z-tubes or rosettes, despite the fact that these orbit types
represent 74\% of all orbits in the galaxy. Notably, inner x-tube orbits are unlikely to interact with the binary despite being radially biased, because they avoid the central region.
Interactions with the SMBH binary cause particles to change their orbits mostly from tube- to box-like orbits, but the other direction also happens. Many box orbits remain boxes after interacting, but usually move to a larger radius. By the final snapshot (1400 Myr) the interacting particles that remain bound are more likely to be on box orbits (72\%) than in the beginning (64\%). Even among the x-tube orbits, interactions favor the radially-biased inner x-tubes rather than the tangential outer x-tubes, that dwindle from 18\% to 1\%. Because of this tendency towards more radially-biased orbits, particles can interact multiple times. Particles that become unbound are predominantly taken from the box orbit reservoir. We cannot exclude that this is also true for the particles ejected between t = 500 Myr and t = 1400 Myr from the 500 Myr tube reservoir. Tube particles might have changed orbital class in between. However, the total fraction is consistent with the strongly kicked tube particles in the smaller 100 Myr intervals.  

\section{Dependence on black hole mass}\label{sec:orbitalfreq}

\subsection{Orbit fractions}
\begin{figure*}
    \centering
    \includegraphics[scale=0.34]{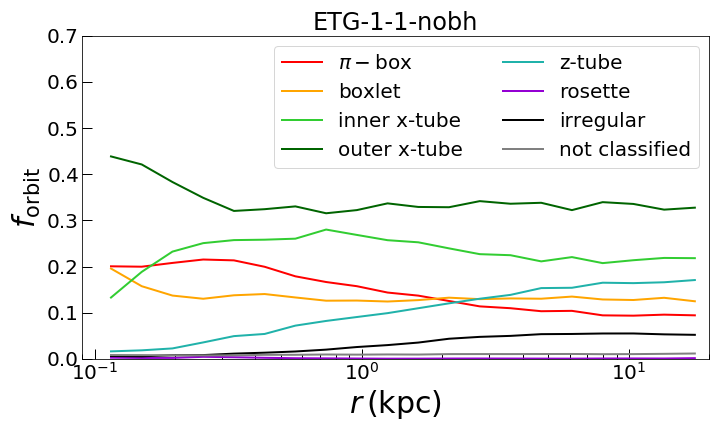}
    \includegraphics[scale=0.34]{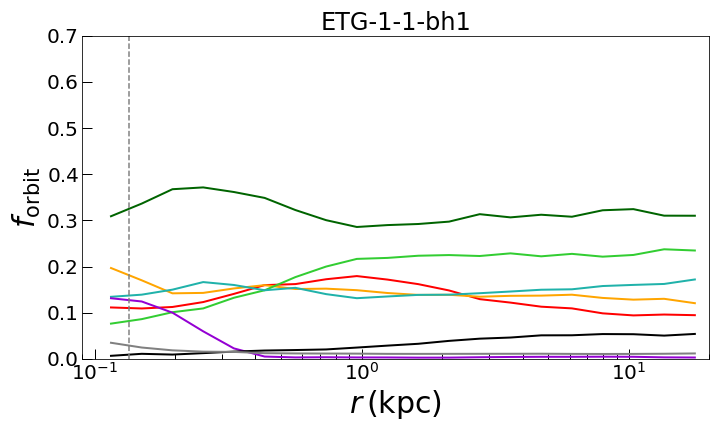}
    \includegraphics[scale=0.34]{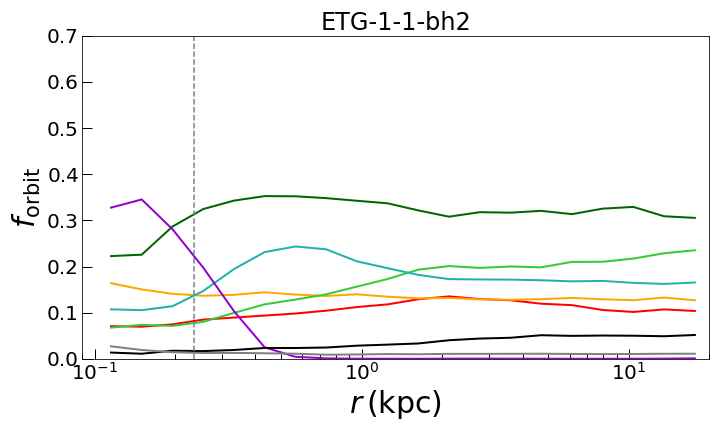}
    \includegraphics[scale=0.34]{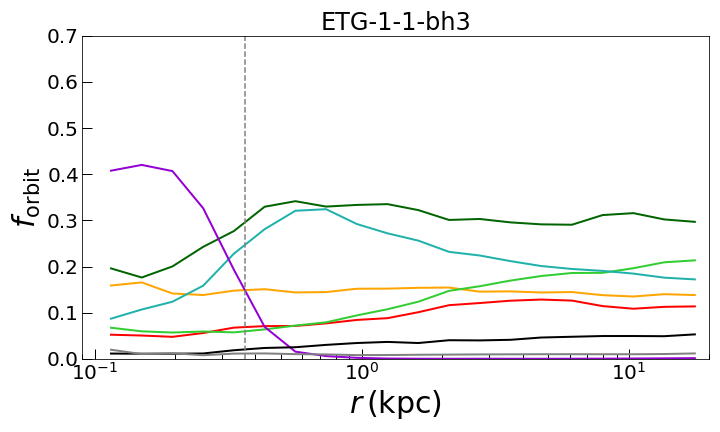}
    \includegraphics[scale=0.34]{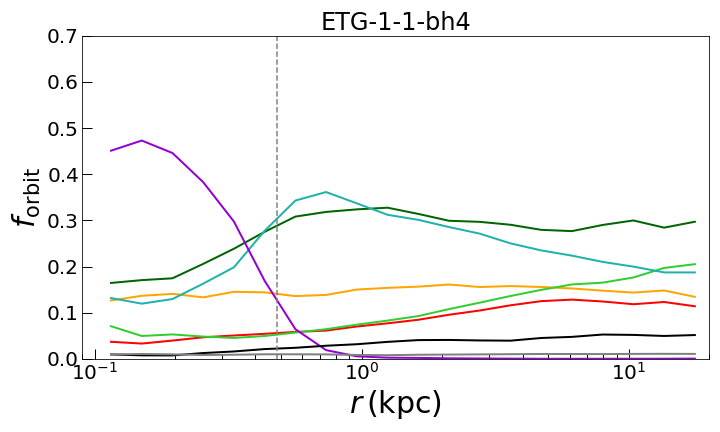}
    \includegraphics[scale=0.34]{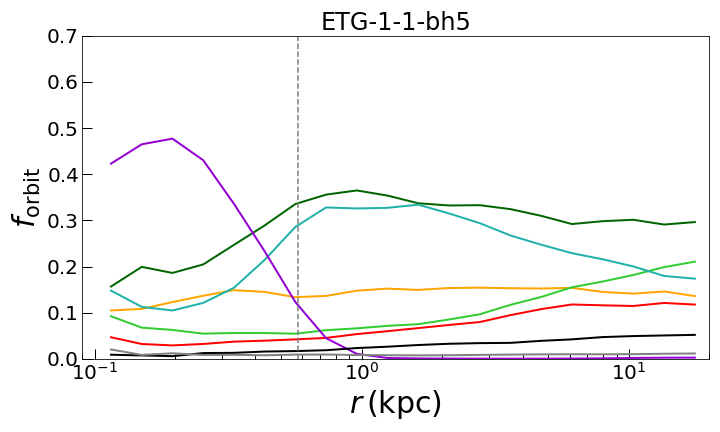}
    \caption[Fraction of the different orbit types in the merger remnants with increasing black hole mass]{Fraction of stellar particles with different orbit classes as a function of radius, for six equal-mass merger simulations without SMBHs and increasing SMBH masses (from top left to bottom right). The prolate remnant without SMBHs (ETG-1-1-nobh) is dominated by x-tubes at all radii. With increasing SMBH mass x-tubes become less common and the fraction of z-tubes increases. At the centre, within the sphere of influence of the SMBH binaries, rosette orbits become the dominant orbits class.  The fraction of \pibox\ orbits decreases. The central region shows the strongest change in its orbital composition due to the more spherical potential generated by the SMBH binaries (see also Fig. \ref{fig:shaperad}). Note the logarithmic scale for the radius on this plot, which highlights the strong changes in the central regions of the merger remnants.}
    \label{fig:gammaradialfreq}
\end{figure*}

\begin{figure}
    \centering
    \includegraphics[width=\columnwidth]{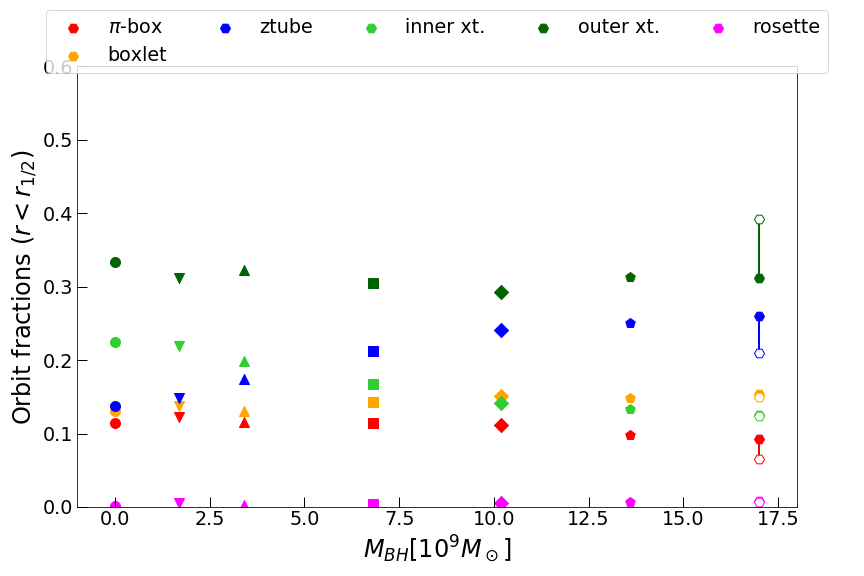}
    \caption[]{Global fractions of orbit classes within the half-mass radius as a function of total SMBH binary mass. All orbits are analysed $\sim$ 1 Gyr after the merger. All merger remnants are prolate on this scale, and as a consequence outer x-tube orbits are the most common family. Inner x-tubes become less common with larger black hole mass, while z-tubes increase correspondingly. Box-like orbit fractions are only weakly affected. In the case of ETG-1-1, with a binary mass of $M_{\mathrm{binary}} = 17 \times 10^9 M_\odot$, (hexagon) we plotted the values at t=300 Myr (empty hexagon) and t=1400 Myr (filled hexagon).}
    \label{fig:allfreqs}
\end{figure}

\begin{figure}
    \centering
    \includegraphics[width=\columnwidth]{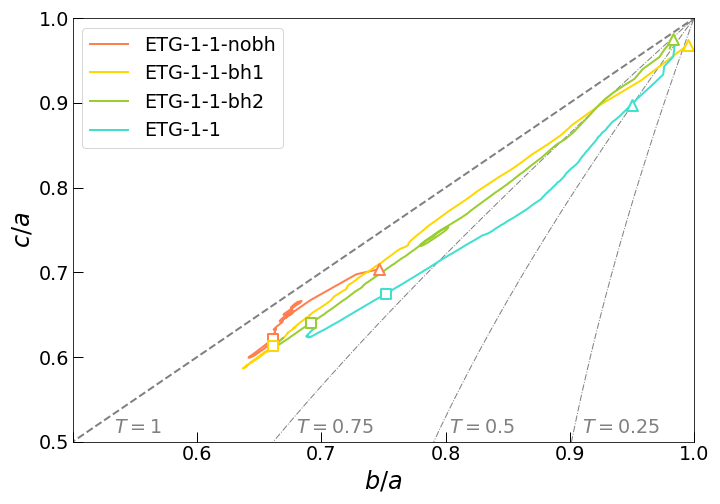} 
    \caption{Axis ratios $c/a$ and $b/a$ of the stellar mass distribution at different radii. The different lines represent different simulations (ETG-1-1-nobh, ETG-1-1-bh1, ETG-1-1-bh2, ETG-1-1) with increasing black hole mass. The triangle marker indicates the value at the core radius, while the square marker indicates the value at the effective radius. The dashed line indicates $b=c$ ($T=1$), while the dot-dashed lines correspond to other constant values of the triaxiality parameter ($T=0.75, 0.5, 0.25$).}
    \label{fig:shaperad}
\end{figure}

In Fig. \ref{fig:gammaradialfreq} we show the fraction of stellar particles in different orbit classes as a function of radius, for the six additional equal-mass merger simulations: ETG-1-1-nobh (no SMBH, top left), ETG-1-1-bh1 ($M_{\mathrm{binary}} = 1.7 \times 10^9 M_\odot$, top right), ETG-1-1-bh2 ($M_{\mathrm{binary}} = 3.4 \times 10^9 M_\odot$, centre left), ETG-1-1-bh3 ($M_{\mathrm{binary}} = 6.8 \times 10^9 M_\odot$, centre right), ETG-1-1-bh4 ($M_{\mathrm{binary}} = 10.2 \times 10^9 M_\odot$, bottom left), ETG-1-1-bh5 ($M_{\mathrm{binary}} = 13.6 \times 10^9 M_\odot$, bottom right). ETG-1-1 with the most massive SMBH binary of $M_{\mathrm{binary}} = 17 \times 10^9 M_\odot$ is already shown Fig. \ref{fig:timeradialfreq}. Again, orbits are classified as \pibox, boxlet, z-tube, x-tube, rosette, irregular, and `not classified' as described in Sec. \ref{sec:orbittypes}. 

Outer x-tubes are the most common orbit family in all simulations and dominate the outer regions of the merger remnants due to the prolate shape of the systems on these spatial scales. Inner x-tubes are very common at small black hole masses, but their fraction quickly decreases with increasing black hole mass, while z-tubes and especially rosette orbits become the most abundant orbit family for the remnants with the most massive SMBH binaries at their centres. This shift is connected to the change in the shape of the central potential of the system, which becomes dominated by the SMBH binaries. Being the remnants of collisionless mergers of spherical systems with a small impact parameter, these galaxies have a prolate shape, which favours x-tube orbits. In the core, however, the mass distribution  is dominated by the SMBH binary, which makes the potential more spherical and allows for more z-tube and rosette orbits within the SMBHs sphere of influence. Rosette orbits represent a small fraction of the overall orbits of the galaxy, but they become more and more prevalent in the centre with increasing black hole mass. 

Box-like orbits become less common in the central regions for systems with larger SMBH masses. In general, this is also connected to the more spherical potential in the centre, which does not support box orbits \citep[see also][]{Barnes1996,Naab2006}. Interestingly however, \pibox\ orbits show the largest difference: in the simulation without black holes they are more common than boxlets, while in the simulation with the most massive black holes they become very rare in the core. In Section \ref{sec:scouring} and particularly in Fig. \ref{fig:timeradialfreq}, we saw that in ETG-1-1 \pibox\ orbits are rare even in the first snapshots after the galactic merger, and actually their number {\it increases} through the slingshot process (Fig. \ref{fig:sankeykicked}), although they move to larger radii. The slingshots mainly suppress boxlet orbits instead. This means that the suppression of \pibox\ orbits induced by the SMBHs must happen very early on. Rather than because of interactions with the binary, they are suppressed by the presence of a point-mass-like  potential in the centre.
Irregular orbits are relatively rare, especially in the central regions, and do not show a strong dependence on black hole mass, but they seem to be slightly more common in the simulation without black holes. 

We summarise the impact of the SMBHs on the global orbit fractions inside the stellar half-mass radius as a function of total SMBH mass of the remnants  in Fig. \ref{fig:allfreqs}. Each simulation is represented with a different marker, and for ETG-1-1 (rightmost value) we indicate the time evolution from t=300 Myr (empty hexagon) to t=1400 Myr (filled hexagon). The stellar half-mass radius of these galaxies is relatively large (about 14 kpc), and the strongest impact of the SMBHs on the orbits is near the centre. Therefore the orbit fractions show only a modest dependence on black hole mass. Outer x-tubes are always the dominant orbit family, with a fraction slightly above 30\% in all remnants. Inner x-tubes are the second most common family at low black hole masses, and they decline with increasing black hole mass, as the galaxy becomes less prolate in the central region. This suggests that inner x-tubes are more susceptible than outer x-tubes to the change of the potential in the core, because of their radial motion. Z-tubes mirror the inner x-tube trend in the opposite direction, growing from less than 15\% in the simulation without black holes to almost 30\% in ETG-1-1. Their fraction increases already at intermediate radii due to the more triaxial shape induced by the SMBHs (see Sec. \ref{sec:shape} and Fig. \ref{fig:shaperad}). The total number of tube orbits (x-tube + z-tube + rosette) is roughly constant at 70\%. Comparing the fraction of tube orbits of ETG-1-1 at different times we see again that x-tubes decrease while z-tubes increase. This effect can only partly be attributed to the black hole binary, as at t = 300 Myr the outer parts of the galaxy were not yet settled. Rosette orbits are negligible at this scale, as they only dominate inside the black hole sphere of influence. Boxlet and \pibox\ orbits make up about 15\% and 10\% of all orbits respectively. Boxlets are roughly constant with black hole mass, while \pibox\ orbits show a slight decline. Interestingly, the fraction of both box orbit types increases with time. The black hole binary in the centre pushes out many box orbits, but they often remain within the half-mass radius, and even many tube/rosette orbits that interact with the binary become box orbits at larger radii (as discussed in Section \ref{sec:scouring}). 

\subsection{Shape of the potential} \label{sec:shape}
The orbital structure of a galaxy is directly linked to the shape of its gravitational potential, which in turn depends on the mass distribution of stars, dark matter and the mass of nuclear supermassive black holes. We can see how the shape of the stellar system is affected by the presence of SMBHs in Fig. \ref{fig:shaperad}. The dashed lines represent different values of the triaxiality parameter $T=\frac{1-(c/a)^2}{1-(b/a)^2}$ ($a$: major axis, $b$: intermediate axis, $c$ minor axis). Each line shows the axis ratios $c/a$ and $b/a$  of the moment of inertia tensor of the enclosed stars going from the half mass radius (squares) to the core radius (triangles) and to the centre of the system.  
The simulation without black holes $ETG-1-1-nobh$ has a prolate stellar body at all radii ( $c/a \sim b/a$, $T \sim 1$). Once a SMBH binary is present the stellar distribution becomes more spherical towards the  centre ( $c/a \sim b/a \sim 1$), while the global shape at $r_{1/2}$ remains prolate. The simulations with the most massive black holes become triaxial ($c/a < b/a, T \sim 0.5$) at intermediate radii. This transition suppresses inner x-tube orbits and favors z-tube orbits in that region (see also Figs. \ref{fig:gammaradialfreq} and \ref{fig:allfreqs}). \newline

\section{Kinematic tomography of a merger remnant}
\label{sec:maps}
\begin{figure*}
    \centering
    \includegraphics[scale=0.34]{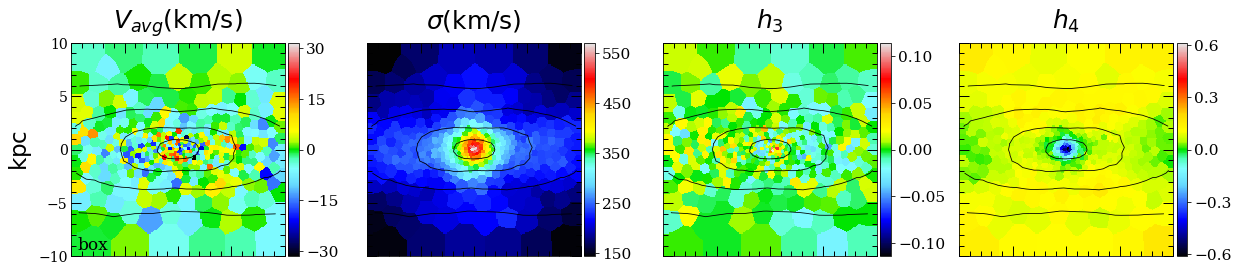}
    \includegraphics[scale=0.34]{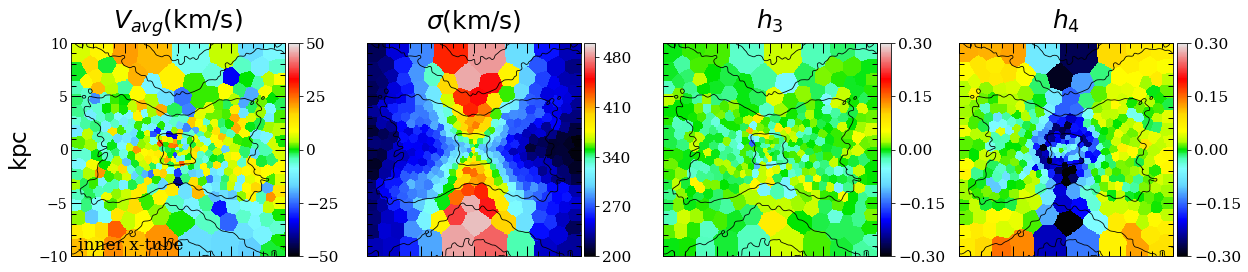}
    \includegraphics[scale=0.34]{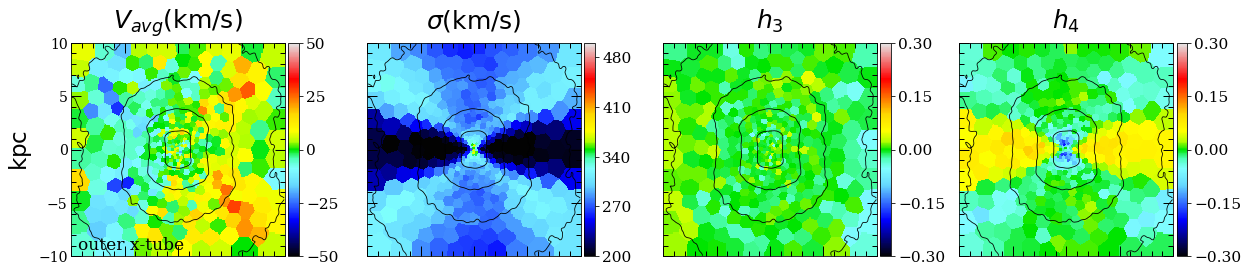}
    \includegraphics[scale=0.34]{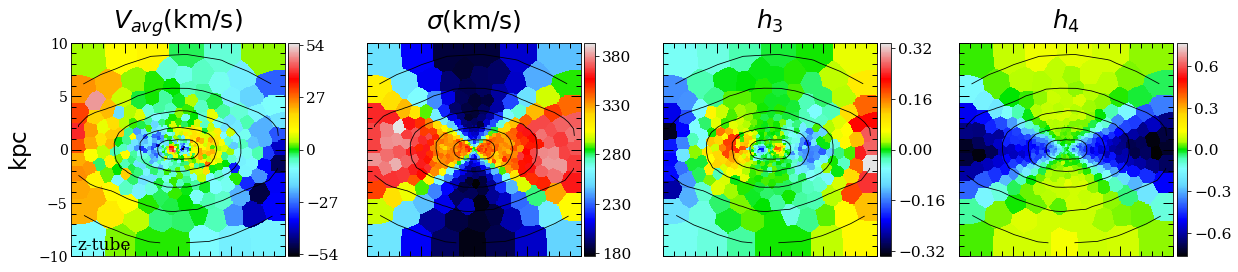}
    \includegraphics[scale=0.34]{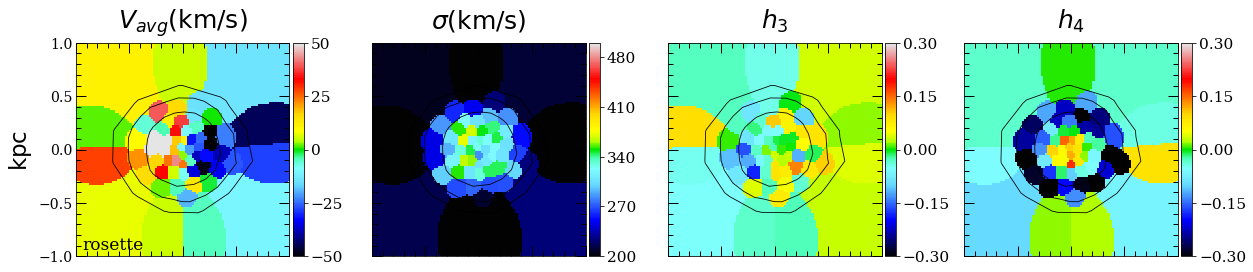}
    \includegraphics[scale=0.34]{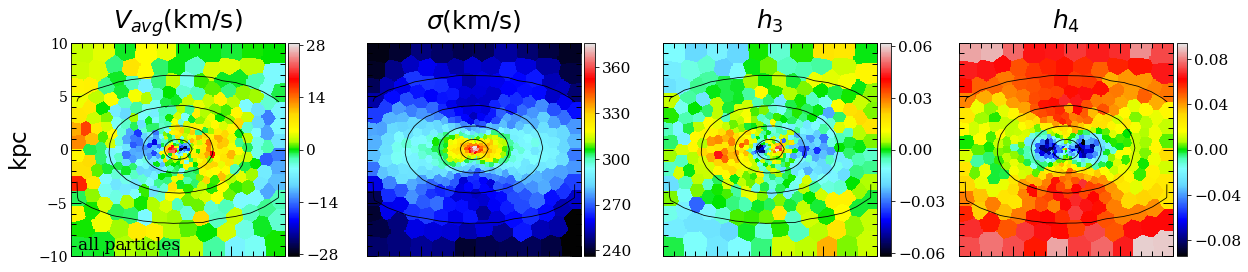}
    \caption[]{Mock two dimensional kinematic maps of ETG-1-1 dissected into the major obit classes. From top to bottom: boxes, inner x-tubes, outer x-tubes, z-tubes, rosettes, and all stars in the galaxy. The four panels of each row show, from left to right, the mean velocity, the velocity dispersion and the third and fourth order Gauss-Hermite moments of the LOSVD, $h_3$ and $h_4$. The black lines are iso-density contours. The map for rosette orbits is zoomed compared to the others, because rosette orbits inhabit only the central regions. The projection is chosen so that the intermediate axis of the galaxy is along the line-of-sight, the minor axis is vertical and the major axis horizontal. The very characteristic features of the different orbit classes are discussed in Sec. \ref{sec:maps}.}
    \label{fig:voroorbit}
\end{figure*}

\begin{figure}
    \centering
    \includegraphics[width=\columnwidth]{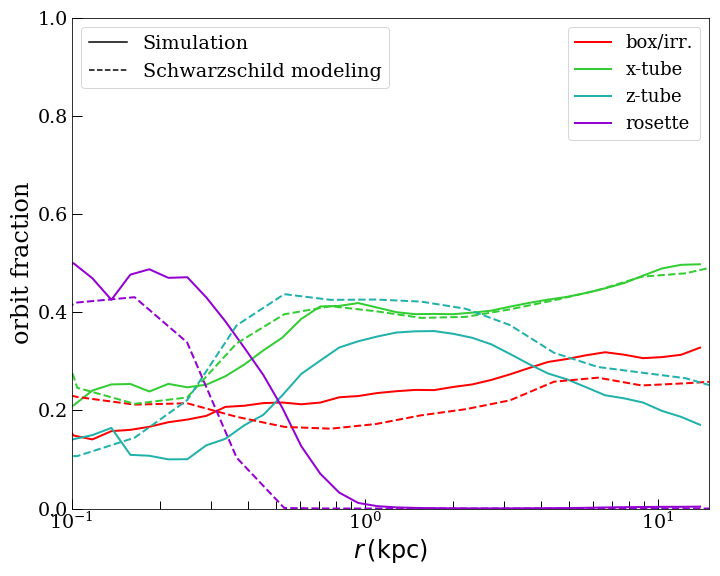}
    \caption[]{Comparison between orbit fractions from our orbit analysis pipeline (solid) and the ones derived from triaxial Schwarzschild modeling (dashed) presented in \citet{Neureiter2021} based on mock-observational kinematic line-of-sight velocity maps (bottom panel of Fig. \ref{sec:maps}) The abundances and their changes with radius agree well between the direct orbit analysis and the Schwarzschild results.}
    \label{fig:sscomp}
\end{figure}
\subsection{Kinematic maps of separated orbital components}
In observations of real galaxies, the dynamical structure of a galaxy has to be inferred from the line-of-sight velocity distribution (LOSVD), often plotted in the form of kinematic maps (e.g. in the ATLAS$^{3D}$ \citep{Cappellari2011}, SAMI \citep{Fogarty2015}, CALIFA \citep{Sanchez2012}, MaNGA \citep{Bundy2015} surveys, or in observations with the MUSE spectrograph, e.g. \citep{Prichard2019}). The superpositions of all orbit classes we have discussed in this paper provide the stellar backbone for these integral field observations.

In Fig. \ref{fig:voroorbit} we plot the global integral field maps and respective contribution of major orbit family  separately for our fiducial  galaxy ETG-1-1.  From top to bottom, we show the maps for box orbits, inner x-tube orbits, outer x-tube orbits, z-tube orbits, rosettes, and the full galaxy. 
For each orbit class, the four panels represent (from left to right) the mean line-of-sight velocity, the velocity dispersion, and the third and fourth order Gauss-Hermite moments \hthree\ and \hfour\ of the LOSVD. 
The irregularly-shaped bins are constructed so that each contains roughly the same stellar mass (in projection), using the Voronoi binning algorithm by \citet{CappellariCopin2003}. As typically done in observations, the LOSVD in each Voronoi bin ({\it spaxel}) is then fitted with a Gauss-Hermite series \citep{vanderMarel1993}, a modified Gaussian that includes terms describing skewness and kurtosis, to obtain the values of mean velocity, velocity dispersion, \hthree, and \hfour. A positive or negative \hthree\ value indicates that the LOSVD has a steep leading wing of particles with negative and positive velocities, respectively. A positive \hfour\ value indicates that the LOSVD is narrow around the mean (i.e. more peaked than the best fitting Gaussian) but has extended high velocity tails (both positive and negative), while a negative \hfour\ value indicates that the LOSVD is relatively flat with weaker tails than a regular Gaussian (or in extreme cases, that the distribution has two peaks). For all the kinematic maps the projection is chosen so that the line-of-sight is aligned with the intermediate axis of the galaxy, the minor axis is vertical and the major axis is horizontal. The approach to mock these observational techniques on simulations is the same as in \citet{Jesseit2007, Naab2014, Roettgers2014,Frigo2019}, where a more detailed description can be found.

In the top row of Fig. \ref{fig:voroorbit} we see that stars on box orbits have very high velocity dispersion in the central region (up to $500 \, \rm km/s$) and low average LOS velocity ($< 50 \, \rm km/s$). This is expected, as these stars have very radial orbits which move fast near the centre of the potential at their pericentre. \hthree\ is almost featureless at all radii. \hfour\ shows strong variations with radius. It is positive at large radii as the LOS velocity distribution is dominated by stars on radial orbits at their apocentre with velocities $\sim 0$ resulting in a peaked LOSVD with extended high-velocity tails. In contrast, the \hfour\ values become very negative towards the centre, where the LOS velocity distribution is relatively flat because of the different orientations (and therefore projected velocity components) of radial orbits at their pericentre. 

Inner and outer x-tube orbits (second and third row) do not show net rotation around the long axis of the galaxy . Their velocity dispersion is very high above and below the midplane and slightly lower in the midplane. This indicates a comparable counter- and co-rotating population of x-tubes. The moderately negative \hfour\ values in the high dispersion region support this. At the centre, where the velocity dispersion is high, \hfour\ becomes very negative ($-0.5$ for inner x-tubes). This indicates a double peaked LOS velocity distribution, with x-tubes rotating in both directions around the long axis of the galaxy. The main difference between the two families lies in the shape of the isophotes and of the high $\sigma$  / low \hfour\ region. Inner x-tubes have a concave shape, and they corotate in the plane perpendicular to the x-axis, while they move more radially elsewhere. Outer x-tubes have a convex shape, and corotate over the entire map except for the midplane.

The z-tubes  (fourth row), however, do show net rotation (up to $\sim 70 \, \rm km/s$) with clearly anti-correlated \hthree\ values. This rotation changes sign in the same pattern as the full kinematic map of the galaxy: once at $0.2$ kpc, once at $1$ kpc, once at $3.5$ kpc. The strength of these patterns gets however reduced by the presence of other orbit types. These patterns are created by the orbit of the SMBHs during the galactic merger. In the first encounter between the two progenitor galaxies, the large amount of ejected mass pulls on the black holes, causing them to invert their orbital angular momentum. Some of the stars that were bound to the black holes before the merger keep an imprint of the merger orbit, which remains visible in the kinematic maps a Gyr after the merger. This process happens twice, explaining the two nested counter-rotating regions. Such counter-rotating patterns have been observed in many real elliptical galaxies \citep{Krajnovic2011, Prichard2019}. The process leading to the features shown here  has been explained in detail in  \citet{Rantala2019}. If we looked at the prograde and retrograde z-tube orbits separately we see that they both rotate very fast (up to $\sim 300 \, \rm km/s$) and more or less balance each other, but the excess of one over the other at different radii causes the counter-rotating features seen in the general kinematic map (bottom row). This kind of superposition resulting in counter-rotating features has also been analysed for observed galaxies (e.g. Fig. 13 of \citealt{vandenBosch2008}). The counter rotating z-tubes show a clear high-dispersion feature along the major axis, again associated with very negative \hfour\ values indicating the double-peaked LOSVD \citep[see e.g.][]{Rix1992}. 

Rosette orbits (fifth row) are shown in a smaller spatial extent (1 kpc) because they only exist within the core of the galaxy. They show net rotation compatible with the z-tube orbits, suggesting that they also conserve angular momentum from the galactic merger.

The global rotation (bottom row of Fig. \ref{fig:voroorbit}) as well as the dominant features in \hthree\ and \hfour\ are mostly generated by the z-tubes, as well as the slightly enhanced dispersion along the major axis at larger radii.   Irregular orbits are rare (see Fig. \ref{fig:allfreqs}), while rosette orbits are clustered around the centre. 

\subsection{A comparison with orbit fractions from Schwarzschild modeling}

For real galaxies observed two-dimensional kinematic maps are used to extract the underlying orbit distributions with Schwarzschild modelling \citep[see e.g.][for an example including box orbits]{vandenBosch2008}. With this technique, the different orbital contributions to the total kinematic map are weighted in order to derive the fractions of different orbit classes at different radii, as well as the mass distribution profile of the galaxy. We have compared our direct orbit classification with results from a novel triaxial Schwarzschild modeling code (SMART, \citealp{Neureiter2021}) that was applied to the stellar density and mock kinematic maps of our fiducial simulation ETG-1-1 (i.e. bottom row of Fig. \ref{fig:voroorbit}). \citet{Neureiter2021} have then followed the identical procedure that is used for a real galaxy and have extracted the radial fraction of rosette, x-tube, box/irregular and z-tube orbits. The orbit analysis in the Schwarzschild models is based on the conservation of the sign of individual components of the total angular momentum vector \citep{Barnes1992}, rather than on a frequency analysis as described in Section \ref{sec:orbitpipeline}. Note also that the code was provided with the 3D stellar density of the galaxy. A full modelling including density profiles obatined from simulated sky images via triaxial deprojections with boxy/discy ellipsoids
\citep{deNicola2020} will be presented in a future work. In Fig. \ref{fig:sscomp} we compare fraction of the different orbit types from our orbit analysis (solid line) with the Schwarzschild modelling result (dashed lines). Box and irregular orbits are grouped together as the Schwarzschild model does not distinguish between them. The Schwarzschild orbit fractions depict the same radial trends as our direct orbit analysis. Rosette orbits dominate the centre ($r < 0.5$ kpc), tubes (both z-tube and x-tube) at the most abundant class at intermediate radii ($0.5 < r < 3$ kpc) with only x-tubes beyond 3 kpc, while box orbits contribute 20-30\% at every radius. The overall good match of the orbit fractions is remarkable. The slightly different ratios of rosette and z-tube orbits can be explained by the fact that these two orbit families share common line-of-sight kinematics (see Fig. \ref{fig:voroorbit}).

The same Schwarzschild model has also been used to estimate the black hole mass at the centre of the system, with an accuracy of a $5 - 10 \%$ depending on the chosen line-of-sight \citep{Neureiter2021}.

\section{Summary}\label{sec:conc6}
In this paper we analyse the dynamical effect of supermassive black hole binaries with varying masses on the structural properties of host galaxies. Our galaxy merger simulations use an accurate integration scheme to account for 3-body interactions of stellar particles with the central SMBH binary. For these simulations we performed a detailed stellar orbit analysis and created mock observational kinematic maps. We find that:
\begin{itemize}

    \item The core in the stellar density profile forms very rapidly on a timescale of tens of Myrs during the final phases of the galactic merger when the sinking (by dynamical friction) black holes form a hard binary. Thereafter the core radius does not increase significantly.  While at core formation the central stellar velocity distribution is still isotropic, it slowly changes towards a very tangentially biased distribution on a timescale of several 100 Myr up to 1 Gyr. This evolution is driven by the removal of particles on radial orbits by slingshot interactions with the central SMBH binary. These interactions move particles to larger radii or kick them out of the galaxy entirely. The slingshot process continues until the SMBHs merge or are ejected from the centre. Therefore the stellar core properties are set in two phases. In the first, short, phase the stellar density core is formed, in the second, extended, phase the core kinematics evolves towards tangential anisotropy.  
    
    \item The evolution of the core stellar velocity anisotropy profile towards more negative values ($\beta \lesssim -0.6$) is driven by the removal of stellar particles from the core, which are on radially-biased orbits with small pericentre distances (\pibox, boxlet). Particles on tube orbits, or on radial orbits that avoid the centre (inner x-tubes) are not as affected, although interactions do still take place.  If kicked particles do not become unbound, they typically end up on box orbits at larger radii. Because of this the overall fraction of box orbits stays roughly constant.
    
    \item The rate of escaping particles is almost constant at $\sim 3 M_\odot \ \rm yr^{-1}$, from after the SMBHs have formed a binary until the end of the simulation (for $\sim$ 1 Gyr). This is likely due to the increasing velocity of the SMBHs as the binary becomes harder, as well as to the movement of the SMBH binary within the stellar core.

    \item All merger remnants, independent of SMBH mass, are dominated by x-tube orbits and have a global prolate shape, caused by the small impact parameter of the merger. With increasing SMBH mass the galactic centres become more spherical and the fraction of x-tubes decreases, especially in the core of the galaxy. Inner x-tubes are especially affected by the change in the potential, because of their larger radial range. At the same time the fraction of z-tube and rosette orbits increases.
    
    \item Box orbits become less common in the core, both with time and with increasing black hole masses. \pibox\ orbits are especially suppressed immediately after the merger took place, because of the change in the shape of the potential (in particular the addition of a point-mass in the centre). The slingshot effect of the black hole binary then further suppresses the fraction of box orbits over time, resulting in the observed velocity anisotropy values. 

    \item We perform a galactic tomography and connect features in the two-dimensional kinematic maps to the responsible orbit classes. Unlike the x-tubes, we find z-tube orbits to show net global rotation, and with high SMBH masses even counter-rotating features. While both prograde and retrograde z-tubes are present at all radii, the slight dominance of one over the other creates the counter-rotating patterns. These coincide with the counter-rotating patterns in the projected kinematic maps of the whole galaxy, and they are connected to the orbital angular momentum flips of the SMBHs orbits during the galactic merger, as studied in \citealp{Rantala2019}. Rosette orbits also show net rotation in the same direction of the central z-tubes.
    
    \item A comparison of our orbits analysis with a novel triaxial Schwarzschild modeling approach based on mock kinematic maps shows a remarkable agreement of the abundance of orbit fractions as a function of radius. This indicates that the orbital structure observed galaxies can be derived at high accuracy. 
    
\end{itemize}
Using a set of idealised simulations we show that supermassive black holes are an important factor for setting the stellar structural and kinematic properties of the centres of elliptical galaxies. The dynamical effects presented here are to be added to the ones of black-hole-powered AGN, which by affecting gas in the galaxy also have a substantial impact on galaxy formation \citep{NaabOstriker2017, Frigo2019}.  In the same way that AGN models are now present in every state-of-the-art cosmological simulation, including accurate black hole interactions will be important to simulate the formation of the most massive galaxies \citep[see e.g.][for a first example]{Mannerkoski2021} and compare with their observed counterparts. Furthermore, with the expanding field of gravitational wave observations, SMBH dynamics might become directly observable in the future through their gravitational wave emissions \citep[e.g.][]{Mannerkoski2019}, opening another window on the role SMBHs play in the formation and dynamics of massive galaxies. 

\section*{Acknowledgments}
TN acknowledges support from the Deutsche Forschungsgemeinschaft (DFG, German Research Foundation) under Germany's Excellence Strategy - EXC-2094 - 390783311 from the DFG Cluster of Excellence ``ORIGINS". P. H. J. acknowledges support by the European Research Council via ERC Consolidator Grant KETJU (no. 818930). The original simulations were carried out at CSC \textregistered \, IT Centre for Science Ltd. in Finland.

\section*{Data availability}
The simulations and data analysis tools underlying this article will
be shared on reasonable request to the corresponding author.

\bibliographystyle{mnras}
\bibliography{mf}

\bsp

\label{lastpage}
\end{document}